\documentclass[12pt]{JHEP3}

\usepackage{amsmath,amssymb,amsfonts,amsthm,mathrsfs}
\usepackage{bm}
\usepackage{graphics}

\newcommand{\be}{\begin{equation}}
\newcommand{\ee}{\end{equation}}
\newcommand{\ba}{\begin{eqnarray}}
\newcommand{\ea}{\end{eqnarray}}
\def\bs{\begin{subequations}}
\def\es{\end{subequations}}
\def\a{\alpha}
\def\b{\beta}
\def\g{\gamma}
\def\e{\epsilon}
\def\ve{\tilde\epsilon}

\def\s{\sigma}
\def\c{{\rm c}}
\def\q{{\rm q}}
\def\vp{\varphi}
\def\dpl{\delta_{\rm Pl}}
\def\cR{{\cal R}}
\def\cH{\mathcal{H}}

\def\cN{{\cal N}}
\def\cV{{\cal V}}

\newcommand{\Eq}[1]{(\ref{#1})}
\def\lp{\ell_{\rm Pl}}

\def\rme{{\rm e}}
\def\rmd{{\rm d}}
\def\rmi{{\rm i}}

\title{Inflationary observables in loop\\ quantum cosmology}

\author{Martin Bojowald$^a$ and Gianluca Calcagni$^b$\\
$^a$ Institute for Gravitation and the Cosmos, The Pennsylvania State University\\
104 Davey Lab, University Park, PA 16802, U.S.A.\\
$^b$ Max Planck Institute for Gravitational Physics (Albert Einstein Institute)\\
Am M\"uhlenberg 1, D-14476 Golm, Germany\\
E-mail: \email{bojowald@gravity.psu.edu}, \email{calcagni@aei.mpg.de}}

\date{November 10, 2010}

\abstract{The full set of cosmological observables coming from linear scalar and
tensor perturbations of loop quantum cosmology is computed in the
presence of inverse-volume corrections. Background inflationary
solutions are found at linear order in the quantum corrections; depending on the values of quantization parameters, they obey an
exact or perturbed power-law expansion in conformal time. The comoving
curvature perturbation is shown to be conserved at large scales, just
as in the classical case. Its associated Mukhanov equation is obtained
and solved. Combined with the results for tensor modes, this yields
the scalar and tensor indices, their running, and the tensor-to-scalar
ratio, which are all first order in the quantum correction. The latter
could be sizable in phenomenological scenarios. Contrary to a pure
minisuperspace parametrization, the lattice refinement parametrization
is in agreement with both anomaly cancellation and our results on
background solutions and linear perturbations. The issue of the choice
of parametrization is also discussed in relation with a possible
superluminal propagation of perturbative modes, and conclusions for
quantum spacetime structure are drawn.}

\preprint{\small Journal of Cosmology and Astroparticle Physics 03 (2011) 032,\qquad\qquad arXiv:1011.2779}


\begin{document}


\section{Introduction}

Loop quantum cosmology (LQC) \cite{LivRev} provides the framework for
implementing several effects seen to arise for the quantum geometry of
loop quantum gravity (LQG) \cite{Rov,ALRev,ThomasRev} in a
cosmological setting. Strict formulations of loop quantum cosmology
are defined on minisuperspace, where one quantizes homogeneous
spacetimes using the methods of loop quantum gravity. The
characteristic effects of discrete spatial geometry then emerge also
in the reduced context, changing the dynamics of expanding universe
models. The dynamics changes in particular at high densities, giving
rise to mechanisms avoiding classical singularities.

Quantum corrections to the classical spacetime structure not only
exist at high densities but remain present, in weaker form, as the
universe expands and dilutes. In such a regime, non-linearities as
well as corrections from loop quantum gravity can be treated
perturbatively in a gauge-invariant way. This has been done for linear
perturbations around spatially flat Friedmann--Robertson--Walker (FRW) models with one particular
class of corrections expected from loop quantum gravity. These
corrections, related to spatial discreteness, arise whenever an
inverse of a certain metric component (or rather, densitized-triad
components) appears in a matter Hamiltonian or in the Hamiltonian
constraint of gravity. Such inverse components are ubiquitous, for
instance in kinetic matter terms, and thus the resulting corrections,
called inverse-volume corrections, are an unavoidable consequence of
loop quantum gravity. Specifically, the corrections are due to the
fact that the quantized densitized triad has a discrete spectrum, with the value zero contained in
the spectrum. Such an operator does not allow the existence of a
densely-defined inverse, but an operator providing the inverse as the
classical limit can nevertheless be defined \cite{QSDV}. When one
departs from the classical regime, however, quantum corrections arise
whose form can be computed in some models and which can be
parametrized sufficiently generally for phenomenological
investigations. In the presence of these corrections, provided they
are small, consistent gauge-invariant cosmological perturbation
equations have been determined \cite{BHKS2}.

Gauge invariance in general relativity and the candidates for its
quantization is intimately related to spacetime structure because
gauge transformations include changes of coordinates. Accordingly, quantum
corrections to gauge-invariant perturbation equations
show how fundamental quantum spacetime effects such as discrete
geometry are reflected, in physical terms, by implications for
cosmological observables. Cosmology then provides an intriguing link
between the fundamental physics of spacetime and phenomenology, as it
will be explored in this paper.

Looking at the issue from the phenomenological side, linearization of
dynamical equations in the presence of inhomogeneous perturbations is
one of the most extensively studied problems in cosmology. With
recent progress in LQC, the context in which such questions can be
addressed has been extended to a new class of quantum-gravity effects
with the aim of making a prediction for early-universe spectra and,
hopefully, constraining the theory. The perturbed equations
contain quantum correction functions and are augmented by
counterterms which guarantee cancellation of anomalies in the
effective constraint algebra \cite{BH1,BH2,BHKS}. These contributions
provide insights into the spacetime structure, modify the
equations of motion and, eventually, imply characteristic
signatures for physical observables.

The perturbed equations of motion for scalar, vector, and tensor modes
have been computed, respectively, in \cite{BHKS2,BH1,BH2}, while the
tensor spectrum and index have been found and explored in
\cite{CMNS2,CH}. To close the set of cosmological consistency relations, it remains to
compute the scalar spectrum and its derived observables. Such is the
\emph{first goal} of this paper: We shall find the Mukhanov equation
of scalar perturbations for the first time and solve it. With exactly
the same procedure, we will rederive the solution of tensor modes with
zero effort. The scalar and tensor spectra and spectral indices will
be obtained together with a consistency relation between the
tensor-to-scalar ratio and the tensor index. In doing so, we will
derive a conservation law for the curvature perturbation, extending
the well-known classical result to the quantum-corrected equations,
and discuss its implications for quantum spacetime structure.

All cosmological observables, including index running and higher-order
quantities, are linearly corrected by quantum terms $\dpl$ whose
magnitude we cannot predict yet because we have no control over the
details of the underlying full quantum theory. The presence of extra
parameters would seem to at best make it possible to place upper
bounds on the quantum corrections $\dpl$ for a given inflationary
potential. For instance, one can arrange to have large-enough quantum
corrections so that the scalar running be sizable. On the other hand,
we can naturally envisage a situation where $\dpl$ is much smaller
than the slow-roll parameters, and therefore completely
negligible. Until better control over the theory is achieved, the
phenomenological consequences of inverse-volume (or other)
corrections will remain unclear but here, as our \emph{second goal},
we highlight the following issue: Different parametrization schemes
will lead to different background solutions and predictions for the
size of $\dpl$. In particular, the usual minisuperspace
parametrization of FRW loop quantum cosmology seems to be incompatible
with anomaly cancellation in inhomogeneous LQC, as well as with the
simplest power-law solutions. Conversely, the lattice refinement
parametrization overcomes all these problems, can predict much larger
quantum corrections, but it might indicate a problem related to
superluminal propagation of the perturbations. We wish to stress all
these features and the importance of further investigating the lattice
refinement parametrization, which requires input from the full theory.

Since anomaly cancellation so far has been shown to occur only in the
quasi-classical regime where inverse-volume corrections and
counterterms are small, we shall concentrate on this case. Therefore
it is not possible to draw comparisons with holonomy-correction
results \cite{CMNS2,Mie1,GB1} or with previous works which
partially fixed the gauge, considered test fields and neglected metric
back-reaction, since they were devoted to the superinflationary regime
of a near-Planckian epoch \cite{TSM,hos04,MN,LQC1,CMNS,SH}.

In section \ref{baeq} we review the LQC background equations of motion
for a flat FRW model and a scalar field, and the
parametrizations arising in minisuperspace quantization and the
lattice refinement phenomenological approach. Background solutions
with de Sitter and exact or perturbed power-law expansion are found in
section \ref{baso}. Section \ref{sper} is completely devoted to scalar
perturbations. After a review and some important updates of the
results of \cite{BHKS2}, we show that the comoving curvature
perturbation is conserved at large scales, as in the classical case
(section \ref{sper2}). The Mukhanov equation for a scalar perturbation
variable is then found and solved in section \ref{sper3}, while the set
of linear cosmological observables (power spectrum, index and index
running) is derived in section \ref{sper4}. The set of observables is
completed in section \ref{tper}, where the same analysis of the previous
section is applied \emph{verbatim} to tensor perturbations. A
discussion of the main achievements of this paper and future
directions can be found in section \ref{disc}.


\section{Spacetime structure and phenomenology}

Loop quantum gravity has provided results that show the discreteness
of spatial quantum geometry: geometrical operators such as those for
area and volume have discrete spectra \cite{AreaVol,Vol2}. Taken by
themselves, these features are not observable because the
corresponding objects, the areas of spatial surfaces or the volumes of
spatial regions, are not gauge invariant. However, these mathematical
properties, derived from the underlying principle of background
independence on which the theory is built, indicate new features of
the quantum representation with important effects for the
dynamics. The volume operator, for instance, enters matter
Hamiltonians \cite{QSDV} and the gravitational Hamiltonian constraint
operator \cite{QSDI}, and thus influences their properties at
the quantum level. In this way, once the dynamical equations are
sufficiently well understood, physical observables are affected and a
potential comparison with observations is made possible.

The dynamics of loop quantum gravity amounts to that of a coupled,
interacting many-body problem in which the elementary constituents are
the fundamental building blocks of space. Such equations governing the
dynamics are difficult to solve exactly, but several crucial effects
visible in them are generic and characteristic; they provide the basis
for phenomenological evaluations. There are two main effects: (i)
inverse-volume corrections and (ii) holonomy corrections due to the
fact that the background-independent quantization used in LQG allows
the representation only of exponentiated curvature components,
gathered as the holonomies of connection variables. Inverse-volume corrections are
currently under much better control in cosmological perturbation
equations, and so they will be our main focus here.

To see implications of those effects in the dynamics, the many-body
Hamiltonians must be analyzed. There is by now a systematic procedure
to do so, based on effective canonical equations to describe
semi-classical dynamics \cite{EffAc,EffCons}. Effective equations in
this context, analogous to low-energy effective actions for
expansions around the ground state of anharmonic systems, are obtained
from expectation values of the Hamiltonian (constraint) operators in a
generic class of semi-classical states. So far, these equations have
been computed and analyzed only in isotropic cosmological models, in
which a solvable system analogous to the harmonic oscillator is
available \cite{BouncePert}. The form of the equations, however, is
more general and can be used also in the presence of inhomogeneous
cosmological perturbations. To avoid bias, one only has to ensure that
correction functions are parametrized sufficiently generally, since
the control over the theory is not yet strong enough to provide unique
predictions for them.

These equations, including parametrized quantum-gravity corrections,
allow important conclusions at the fundamental and phenomenological
levels by a clear line of arguments. First, the structure of spatial
geometry changes according to LQG, affecting the form of the
constraints generating gauge transformations. Secondly, in the presence of
quantum corrections the algebra under Poisson brackets (or
commutators when quantized) of the constraints as gauge generators is
modified, amounting to a different realization of the classical
transformations of spacetime and the underlying gauge behaviour. This
abstract feature has several consequences. For instance,
quantum-geometry corrections cannot be implemented by any
higher-curvature effective action because those corrections would not
change the classical constraint algebra and the underlying notion of
gauge and covariance. An effective action that can include all
effects of LQG must be of a more general form, for instance allowing
for non-commutative geometry.  Moreover, gauge transformations
belonging to a deformed algebra no longer correspond to ordinary
coordinate transformations on a manifold. Thus, effective line
elements may be questionable because the transformations of metric
components and coordinate differentials collected in $\rmd s^2$ no longer
match to make the line element invariant. In such a context, physical
information can be gained only from gauge-invariant variables that
take into account the new gauge structure. With these additional
effects, there is a chance that quantum gravity corrections may be stronger than usually
expected, for instance, from naive arguments based on the size
of higher-curvature corrections.

One possible new phenomenon is of particular interest. When the
constraint algebra is deformed, the Bianchi identity or the
conservation equation for stress-energy is modified. There are still
analogous identities if the deformation is consistent and
anomaly-free, but they may refer to different quantities than in the
classical case. Then, the curvature perturbation is no longer
guaranteed to be conserved, a possibility which has already been raised \cite{BHKS2,InhomEvolve}. If the curvature perturbation is no
longer conserved, on the other hand, magnification effects for modes
outside the Hubble radius can be expected. Even though deviations from
conservation given by quantum gravity were small at any given
(sufficiently late) time, expected to be determined by the tiny ratio
of the Planck length by the Hubble radius, the lever arm of
non-conservation during long times between horizon exit during
inflation and re-entry might magnify those tiny effects. In this way,
a tight link is obtained between fundamental spacetime structure and
cosmological phenomenology.

Here we will provide results in both directions, fundamental
physics as well as phenomenology. In particular, we will demonstrate
that a subtle cancellation in the anomaly-free correction functions of
LQC does make the curvature perturbation conserved in spite of the
non-trivial deformation of the constraint algebra, a feature which has
not been noticed before. As a consequence, effective linear
perturbations of Friedmann--Robertson--Walker geometries can be
meaningfully constructed even in the presence of a deformed gauge
structure.  Corrections to standard perturbation equations then follow
naturally.


\section{Background equations and parametrizations}\label{baeq}

To begin, we write down the LQC effective equations of motion for an
FRW background $\rmd s^2=a^2(\tau)(-\rmd\tau^2+\rmd x^i\rmd x_i)$ in conformal time
$\tau$ (see, e.g., the review in \cite{CH} for a detailed derivation of
these results). We shall ignore holonomy corrections, which have not
yet been considered in the perturbed dynamics. Notice, however, that
these contributions in some parametrizations dominate over
inverse-volume corrections as far as tensor modes are concerned
\cite{CH,GB1}; we will further discuss this point in the final
section. For a scalar field $\vp$ with potential $V$, the effective
Friedmann and Klein--Gordon equations read
\be\label{frw}
\cH^2=\frac{8\pi G}{3}\,\a\left[\frac{{\vp'}^2}{2\nu}+pV(\vp)\right]
\ee 
and 
\be\label{kg} 
\vp''+2\cH\left(1-\frac{\rmd\ln\nu}{\rmd\ln
p}\right)\vp'+\nu p V_{,\vp}=0\,, 
\ee 
where $G$ is the gravitational constant, $\cH\equiv a'/a$ is the
Hubble parameter, primes denote derivatives with respect to conformal
time, and $p=a^2$ (in comoving volume units) is the triad variable in
minisuperspace. (Triad variables can take both signs depending on
the orientation of space. Here we assume $p$ to be positive without
loss of generality for effective equations.) From equations (\ref{frw})
and (\ref{kg}) one obtains the Raychaudhuri equation
\begin{equation} \label{Ray}
 \cH'= \left(1+\frac{\rmd\ln\alpha}{\rmd\ln p}\right)\cH^2- 4\pi G 
 \frac{\alpha}{\nu} \left(1-\frac{1}{3}\frac{\rmd\ln\nu}{\rmd\ln p}\right)
 {\vp'}^2\,.
\end{equation}

In these equations,
\ba \a
&\approx& 1+\a_0\dpl,\label{an}\\ 
\nu &\approx& 1+\nu_0\dpl,\label{cn}
\ea 
where 
\be\label{dp} \dpl\equiv
\left(\frac{p_{\rm Pl}}{p}\right)^{\frac{\s}{2}}=\left(\frac{a_{\rm Pl}}{a}\right)^\s\,,
\ee 
and $\s$ and $p_{\rm Pl}=a_{\rm Pl}^2$ are constant.\footnote{We put a subscript
`Pl' in the definition \Eq{dp} in order to avoid confusion with
perturbations such as $\delta\vp$. However, the equations below do not
rely on any particular characteristic scale $a_{\rm Pl}$, which may differ
from the Planck length.} We will often need to switch from $p$ to conformal derivatives via 
\be {}' = 2\cH\frac{\rmd}{\rmd\ln p} 
\ee 
and the formul\ae\ 
\be \dpl'=-\s\cH\dpl\,,\qquad
\dpl''=\s\cH^2(\s-1+\e)\dpl\,,
\ee
with $\e=1-{\cal H}'/{\cal H}$ introduced as a slow-roll parameter below.

Functional forms of $\alpha(p)$ and $\nu(p)$ are fully computable in
general form from operators in exactly isotropic models
\cite{InvScale} and for regular lattice states in the presence of
inhomogeneities \cite{QuantCorrPert}, with parametrizations of
quantization ambiguities affecting the values of $\alpha_0$, $\nu_0$
and $\sigma$ \cite{Ambig,ICGC}. However, only the expanded forms
(\ref{an}) and (\ref{cn}) are needed in the perturbative regime
considered here. From these explicit calculations of inverse-volume
operators and their spectra, one can derive further properties
characteristic of loop quantum gravity. In particular, correction
functions implementing inverse-volume corrections, when evaluated at
large values of the densitized triad or the scale factor in a nearly
isotropic geometry, approach the classical value always from
above. This consequence, which is a robust feature under quantization
ambiguities and will turn out to be important later, implies that the
coefficients $\a_0$ and $\nu_0$ introduced in the parametrizations
used here must be positive.


\subsection{Parametrizations}

The above equations are derived in a minisuperspace Hamiltonian
formalism where the super-Hamiltonian (the only non-trivial constraint
on homogeneous backgrounds) is first symmetry-reduced, and then
quantized with LQG techniques. The resulting equations then constitute
partial effective equations, which means that they capture the
behaviour of expectation values of observables in semi-classical states,
but without taking all quantum corrections into account. In
particular, quantum back-reaction by fluctuations and the holonomy
corrections of loop quantum gravity are not included at the present
stage. The LQG techniques consist in a choice of canonical variables
and operator ordering which make the final result quite different with
respect to the traditional minisuperspace Wheeler--DeWitt quantum
cosmology. At the semi-classical level, the main difference is the
presence of correction functions \Eq{an} and \Eq{cn}. The constants
$\a_0$, $\nu_0$ and $\s$ will enter the cosmological observables and
it is important to set their value range beforehand. This range
strongly depends on the physical interpretation of the model. We can
identify two views on the issue, one purely homogeneous and isotropic
and the other associated with the natural presence of inhomogeneities.

\subsubsection{Minisuperspace parametrization}\label{misup}

On an ideal FRW background, open and flat universes have infinite
spatial volume and the super-Hamiltonian constraint is formally ill
defined because it entails a divergent integration of a spatially
constant quantity over a comoving spatial slice $\Sigma$,
\be\nonumber
\int_\Sigma \rmd^3 x=+\infty\,.
\ee
To make the integral finite, it is customary to define the constraint
on a freely chosen finite region of size $\cV=a^3\cV_0$, where $\cV_0$ is the corresponding comoving volume:
\be\nonumber
\int_\Sigma \rmd^3 x\to\int_{\Sigma(\cV_0)} \rmd^3 x=\cV_0<+\infty\,.
\ee
The volume appears in the correction function \Eq{dp} as $\dpl\sim
a^{-\s}\sim \cV^{-\s/3}$. To make $\dpl$ adimensional, one can use
the Planck length $\lp$ to write
\be\label{dpmi} 
\dpl \sim
\left(\frac{\lp^3}{\cV_0}\right)^{\frac{\s}{3}} a^{-\s}\,.  
\ee 
To specify the coefficient further, one sometimes introduces the area gap
$\Delta_{\rm Pl}\equiv 2\sqrt{3}\pi\gamma\lp^2$, where $\gamma$ is the
Barbero--Immirzi parameter.\footnote{The use of twice this value
according to recent findings may be better justified \cite{AsW} but the
resulting change in the values of $\a_0$ and $\nu_0$ is not relevant
for what follows.}  A detailed calculation then shows that the
constant coefficients $\a_0$ and $\nu_0$ are \cite{CH}
\be
\a_0=\frac{(3q-\s)(6q-\s)}{2^2 3^4}\left(\frac{\Delta_{\rm Pl}}{p_{\rm Pl}}\right)^2,\qquad \nu_0= \frac{\s(2-l)}{54}\left(\frac{\Delta_{\rm Pl}}{p_{\rm Pl}}\right)^2\,.\label{ten}
\ee
They depend on two sets of ambiguities, one ($1/2\leq l<1$ and
$1/3\leq q< 2/3$) related to different ways of quantizing the
classical Hamiltonian\footnote{The different interval for $q$ with
respect to the one given in \cite{CH} stems from the same argument in
the full theory which constrains the range of $l$
\cite{QSDV,boj12}. We set the natural value of $q$ to be $1/2$ rather
than 1 \cite{APS} in equation \Eq{natu}.} and the other ($\s$) depending on
which geometrical minisuperspace variable has an equidistant stepsize
in the dynamics: in terms of the triad variable $p$, $p^{\sigma/2}$ is
equidistant if inverse-volume corrections with exponent $\sigma$ in
$\Delta_{\rm Pl}$ appear. More physically, this parameter is related
to how the number of plaquettes of an underlying discrete state
changes with respect to the volume as the universe expands. The latter
is a phenomenological prescription for the area of holonomy
plaquettes, but ideally it should be an input from the full theory
\cite{bo609}.

In the minisuperspace context, a natural choice of these parameters is 
\be\label{natu} 
\s=6\,,\qquad
l=\tfrac34\,,\qquad q=\tfrac12\,, 
\ee 
so that, assuming
$p_{\rm Pl}=\Delta_{\rm Pl}$ \cite{APS}, one has 
\be 
\a_0=\tfrac1{24}\approx
0.04\,,\qquad \nu_0=\tfrac5{36}\approx 0.14\,.\label{nulc0} 
\ee 
Notice that $\s=6$, motivated by holonomy corrections not becoming
large at small curvature, corresponds to the so-called `improved
quantization scheme' \cite{APS}, a name which applies also to values
of $l$ and $q$ different from equation \Eq{natu}. A specific example
provided in \cite{APS}, for instance, is $q=1$ such that
$\a_0=0$. Note, however, that the exact value $\a_0=0$ is obtained in
this case only due to a spurious cancellation in isotropic settings;
it seems to suggest negligible inverse-volume corrections but is
unreliable compared with more general derivations. For phenomenology
at the current level of precision, the most significant parameter is
$\s$, which is not as much affected by different choices of the
minisuperspace scheme.

Since $\delta_{\rm Pl}$ is ${\cal V}_0$-dependent, inverse-volume
corrections cannot strictly be made sense of in a pure minisuperspace
treatment. Inverse-volume corrections, as used here, have never been
derived fully consistently in this context owing to the
$\cV$-dependence. In particular, as discussed in
\cite{SchwarzN,Consistent}, while the improved scheme does take into
account refinement for holonomies in an ad-hoc manner, it ignores
these effects for inverse-volume corrections. This failure to
represent inverse-volume effects, which are crucial for well-defined
Hamiltonians in loop quantum gravity, presents a serious limitation of
pure minisuperspace models which can be overcome only by bringing in
further ingredients to take into account the behaviour of
inhomogeneities, as indicated in what follows. Precise derivations
become more complicated in this situation, but by a combination with
input from phenomenology one can obtain valuable restrictions on the
possible realizations.

We wish to make a further comment on this issue. In the context of
inflation, one has a set of observables given by the anisotropy
spectra and their derivatives; all these quantities contain parameters of the inverse volume corrections. Since they are all
evaluated at horizon crossing, the comoving fiducial volume therein
(implicitly conceived as greater than the causal region $\sim
\cH^{-3}$) can be naturally set to be the Hubble volume. If one
maintained the minisuperspace parametrization also in the presence of
perturbations, the conclusion would be that the fiducial volume
problem is less severe than expected. However, the minisuperspace
setting pertains only to exactly isotropic models, and this solution
of the problem is at best incomplete. To get a clearer picture, we
should include inhomogeneities already at the fundamental level. The
following argument shows how to do so qualitatively.

\subsubsection{Lattice refinement parametrization}

The chosen volume $\cV$ is a purely mathematical object which
should not appear in physical observables, but it does appear in
equation \Eq{dpmi}. Since $\dpl$ will enter the observables, we might face
a problem. To make the situation better behaved, we introduce generic
effects of inhomogeneities. One example for doing so is the lattice
parametrization discussed in \cite{CH} which, as one implication,
extends the range of the parameters. Then,
one has a large range of $\s$,
\be
4<\s\leq 6\,,\label{ci}
\ee
the value of the improved minisuperspace quantization scheme being
included. When $p_{\rm Pl}=\Delta_{\rm Pl}$, for instance, varying over the
range of $\s$ (equation \Eq{ci}), $l$ and $q$ (as specified below
equation \Eq{ten}),
\ba
0<&\a_0&\leq\tfrac5{81}\approx 0.06\,,\label{alc}\\
0.07\approx \tfrac2{27}<&\nu_0&<\tfrac16\approx 0.17\,.\label{nulc}
\ea

Instead of repeating the arguments leading to equation \Eq{ci}, we consider
an alternative lattice parametrization where one uses the `patch'
volume of an underlying discrete state in correction functions, rather
than the much larger volume $\cV$ \cite{Consistent}. In this
parametrization, motivated by key aspects of discrete spacetime
dynamics, the ranges of parameters change more significantly than in
(\ref{ci}), with important consequences for phenomenology. 
Now, corrections refer to the patch size 
\be
v\equiv \frac{\cV}{\cN}\,,
\ee
with $\cN$, the main input from quantum gravity, the number of
discrete patches in $\cV$. By construction, $v$ is independent of the
size of the region, since both $\cV_0$ and ${\cal N}$ scale
in the same way when the size of the region is changed. Physical
predictions should not feature the region one chooses unless one is
specifically asking region-dependent questions (such as: What is the
number of vertices in a given volume?).

Given the available parameters and their dimensions, the
leading-order quantum correction in $\alpha$ and $\nu$ is then of the
form
\be\label{m} 
\dpl= \left(\frac{\lp^3}{v}\right)^{\frac{m}{3}}=
\left(\lp^3\frac{{\cal N}}{\cV}\right)^{\frac{m}{3}}, 
\ee 
where $m>0$ is a
positive integer parameter. It determines the power by which $p$
appears in leading corrections of an expansion of inverse-volume
correction functions.  Primarily, the correction functions
$\alpha$ and $\nu$, and thus $\dpl$ as well, depend on flux values,
corresponding to $p$ for the isotropic background. Since $p$ changes
sign under orientation reversal but the operators are parity
invariant, only even powers of $p$ can appear, giving $m=4$ as the
smallest value. At this stage of development of the full theory, it is
not entirely clear that general correction functions depend only on
the fluxes (rather than, e.g., also on the eigenvalues of more
complicated volume operators; for properties of their spectra see
\cite{VolSpecI,VolSpecII}). Therefore we set $m\geq 4$.

A time-dependent ${\cN}(t)$ corresponds to the dynamical
`lattice-refinement' behaviour \cite{bo609}. For some stretches of
time, one can choose to use the scale factor $a$ as the time variable
and represent ${\cal N}(a)$ as a power law
\be
\cN=\cN_0a^{-6x}\,,
\ee
where $\cN_0$ is some (coordinate and $\cV_0$-dependent) parameter and
the power $x$ describes different qualitative behaviours of changing lattices. Overall, we have
\be\label{13}
\dpl= \left(\lp^3\frac{\cN_0}{\cV_0}\right)^{\frac{m}{3}} a^{-(2x+1)m}\,.
\ee
This equation cannot be obtained in a pure minisuperspace setting
(e.g., \cite{CH}) where only one parameter $\sigma$ enters as in
(\ref{dpmi}). The presence of an extra parameter, compared to the
minisuperspace parametrization, may appear as a disadvantage, but we
will see later that it is required for being able to match with
phenomenology; otherwise the model would be ruled out.

In the lattice-refinement derivation, the parameter $a$ plays
two roles, one as a dynamical geometric quantity and the other as
internal time. While writing down the semi-classical Hamiltonian with
inverse-volume (and holonomy) corrections, one is at a non-dynamical
quantum-geometric level. Then, internal time is taken at a fixed value
but the geometry still varies on the whole phase space. In this
setting, we must keep $\cN$ fixed while formulating the constraint as
a composite operator. The net result is the Hamiltonian constraint
operator of the basic formulation of loop quantum cosmology
\cite{cosmoII,IsoCosmo} not taking into account any refinement,
corresponding to $x=0$ and $m=\s$. However, when one solves the
constraint or uses it for effective equations, one has to bring in the
dynamical nature of $\cN$ from an underlying full state. This is the
motivation for promoting $\cN$ to a time-dependent quantity, a
step which captures operator as well as state properties of the
effective dynamics. Its parametrization as a power law of the scale
factor is simply a way to encode the qualitative (yet robust, see
below) phenomenology of the theory. The general viewpoint is similar
to mean-field approximations which model effects of underlying degrees
of freedom by a single, physically motivated function.

Comparing with the earlier minisuperspace parametrization, equation \Eq{13}
gives $\s=(2x+1)m$ as far as the $a$-dependence is concerned. The
number of vertices $\cN$ must not decrease with the volume, so $x\leq
0$; it is constant for $x=0$. Also, $v=\cV/\cN\sim a^{3(1+2x)}$ is the
elementary geometry as determined by the state; in a discrete
geometrical setting, this quantity has a lower non-zero bound
which requires $-1/2\leq x\leq 0$. In particular, for $x=-1/2$ we have
a constant patch volume, corresponding to what is assumed in the
improved minisuperspace quantization scheme \cite{APS}. In contrast
with the minisuperspace parametrization \Eq{natu}, in the effective
parametrization of equation \Eq{13} we have $\s=0$ in this case. Thus, even
for $x=-1/2$ is the parametrization new and different from the
minisuperspace representation, overcoming the problem of representing
inverse-volume effects in a pure minisuperspace treatment.

To summarize the general lattice-refinement scheme, $\s=(2x+1)m$ is a
time-independent or slowly changing parameter\footnote{The
  creation or subdivision of new cells in a discrete state depends on
  the spatial geometry and can thus be considered as changing more
  slowly than other processes in an expanding universe. On large time
  scales, the parameter $\sigma$ may change, distinguishing different
  microscopic epochs in the history of the universe.} given by the
reduction from the full theory and with range 
\be\label{newc} 
\s\geq 0\,.  
\ee 
Assuming this range will be of utmost importance for
justifying the validity of the cosmological perturbation spectra. In
fact, we shall find that $\s$ must be small in order
for the spectra to be almost scale invariant, a range of values that
cannot occur for either the minisuperspace parametrization or the
first lattice parametrization \Eq{ci}. Responsible for the better
matching is the new parameter $x$, while $m$ alone (or $x=0$) would
give a range similar to the minisuperspace parametrization.

Note that, in principle, $\s$ may be different in $\a$ and $\nu$ for
an inhomogeneous model. However, here we assume that the background
equations \Eq{an} and \Eq{cn} with the same $\delta_{\rm Pl}$ are
valid also in the perturbed case. Not only is this choice natural
whenever background quantities are considered, but it is also crucial
for several simplifications to follow.

Before moving on, a remark is in order. The patches of volume $v$ find
a most natural classical analog in inhomogeneous cosmologies, in
particular within the separate universe picture \cite{WMLL}. For
quantum corrections, the regions of size $v$ are provided by an
underlying discrete state and thus correspond to quantum degrees of
freedom absent classically. However, the discrete nature of the state
implies that inhomogeneities are unavoidable and no perfectly
homogeneous geometry can exist. Given these inhomogeneities and their
scale provided by the state, one can reinterpret them in a classical
context, making use of the separate universe picture. There, the volume $\cV$ can be regarded as a region of the universe
where inhomogeneities are non-zero but small. This region is coarse
grained into smaller regions of volume $v$, each centered at some
point ${\bf x}$, wherein the universe is FRW and described by a
`local' scale factor $a(t,{\bf x})=a_{\bf x}(t)$. The difference
between scale factors separated by the typical perturbation wavelength
$|{\bf x}'-{\bf x}|\sim \lambda\ll \cV^{1/3}$ defines a spatial
gradient interpreted as a metric perturbation. In a perfectly
homogeneous context, $v\sim \cV$ and there is no sensible notion of
cell subdivision of $\cV$; this is tantamount to stating that only the
fiducial volume will enter the quantum corrections and the
observables, $\cN=\cN_0$. On the other hand, in an inhomogeneous
universe the quantity $v$ carries a time dependence which, in turn,
translates into a momentum dependence. The details of the cell
subdivision (number of cells per unit volume) are intimately related
to the structure of the small perturbations and their
spectrum. Thus, lattice refinement is better suitable in the
cosmological perturbation analysis. As long as perturbations are
linear and almost scale invariant, the size of volume within which the
study is conducted is totally irrelevant.

\subsubsection{Parameter estimates}\label{s:Estim}

Since the lattice refinement picture is phenomenological, presently we
are unable to determine the quantum correction $\dpl$ from first
principles and, at this stage, the latter is regarded as a free
parameter which can be constrained by experiments. In the
minisuperspace and lattice parametrization \Eq{ci}, on the other hand,
there may be an argument which estimates the magnitude of $\dpl$
heuristically. In fact, in those parametrizations
$\dpl=(\delta_0/\cV)^{\s/3}$, where $\delta_0$ is some constant volume
and $\cV$ is the fiducial volume; for lack of better knowledge, one
often assumes $\delta_0\sim \lp^3$. All inflationary observables are
evaluated at horizon crossing, so the volume $\cV$ is very naturally
fixed by the size of the Hubble horizon at that moment (denoted with a $*$) \cite{CH}:
\be\label{boh} 
\cV\sim H^{-3}_*\,,
\ee 
where $H=\dot a/a=\cH/a$. (This equation is invariant
under isotropic rescalings of the coordinates. In terms of $a_{\rm Pl}$
introduced via $\dpl=(a_{\rm Pl}/a)^{\sigma}$ in the minisuperspace
parametrization, we may write $a_{\rm Pl}/a=\lp H_*$.)  The point here is
that so far $\cV$ has been arbitrary, the only requirement being that
it contains the Hubble region at any given moment. Provided $\delta_0$
is fixed \emph{a priori}, this equation fixes $\cV$ once and for all
because one is not at the liberty of changing the numerical factor in
\Eq{boh}, which is ${\rm O}(1)$. Slightly different definitions of the
Hubble horizon differ only for ${\rm O}(1)$ coefficients, which do not
affect the discussion qualitatively (on the other hand, ${\rm O}(10)$ or
${\rm O}(0.1)$ coefficients are unacceptable because the observables here
are defined at, not before or after, horizon crossing).

To estimate $\dpl$ during inflation, we could take the
grand-unification scale $H_*\sim 10^{14} \div 10^{17}~{\rm GeV}$. As
$\dpl=(\lp H_*)^\s$ and $4<\s\leq 6$, one has the upper bound
$\dpl\lesssim {\rm O}(10^{-8})$. Typical prefactors in observable quantities
may even carry an extra ${\rm O}(10^{-1})$ suppression, as we will find, so
none of the inverse volume corrections with these choices are
observable, even in the scalar running, as the slow-roll parameters
are ${\rm O}(10^{-2})$. In comparison, the inflationary tensor index gets an
extra contribution $\delta_{\rm hol}\propto(\lp H_*)^2\lesssim
{\rm O}(10^{-4})$ from holonomy corrections \cite{GB1}, which dominate
over $\dpl$. Holonomy corrections in the perturbed scalar sector has
never been computed, but we expect a similar hierarchy of scales.

Unfortunately, these estimates rely on a particular choice for
$\delta_0$ and the argument cannot be regarded as robust. The size of
this dimensionful constant strongly depends on the underlying theory,
and a change in magnitude of $\delta_0$ would modify the above
results. Moreover, we will see that the minisuperspace
parametrizations are not compatible with requirements on background
solutions combined with anomaly freedom. The estimate of $\dpl$ could
thus at best be used as an external input for the lattice-refinement
parametrization. Then, once the scale of inverse-volume corrections is
fixed, a further consistency condition must be satisfied because
inverse-triad corrections and holonomy corrections are determined by
the same parameter that specifies the underlying discreteness scale
\cite{Consistent}. To evaluate this condition, we use the alternative
form of $\dpl=(\lp^3/v)^{m/3}= \lp^4/v^{4/3}$ for $m=4$.  The
underlying discreteness scale, as a distance parameter, is then the
linear dimension of patches, $L=v^{1/3}= \lp/\dpl^{1/4}= \lp/(\lp
H_*)^{\sigma/4}$. On the other hand, the strength of holonomy
corrections can be expressed in terms of the critical density
$\rho_{\rm crit}= 3/(8\pi G\gamma^2L^2)$;\footnote{As an explicit
calculation shows, the true critical density is actually $\a\rho_{\rm
crit}$ \cite{CH}. Setting $q=1$ and $\s=6$ in the lattice
parametrization one gets $\a=1$, as in \cite{APS}. In general $\a$
does appear, but the arguments in this discussion are qualitative and
we can ignore this issue.} holonomy corrections are weak when the
matter density satisfies $\rho\ll \rho_{\rm crit}$. With $L$ as
assumed here,
\begin{equation} 
 \rho_{\rm crit}=\frac{3 H_*^2 (\lp^2H_*^2)^{\sigma/4-1}}{8\pi G\gamma^2}
\sim \rho_*\left(\frac{\rho_*}{\rho_{\rm Pl}}\right)^{\sigma/4-1}\,,
\end{equation}
with the matter density $\rho_*$ at the time of horizon
crossing. Thus, for $\rho_{\rm crit}\gg \rho_*$ to ensure that
holonomy corrections do not significantly alter the classical
behaviour at horizon crossing, we must require $\sigma<4$. Then, $\dpl$
becomes larger than estimated above. For a critical density of
Planckian size, which is often desired so as to have strong
quantum-gravity corrections only in the Planckian regime, $\sigma$
must be close to zero. For such values, $\dpl\sim {\rm O}(1)$ with the above
estimate of $\delta_0$, clearly
dominating holonomy corrections at horizon crossing. (For $\sigma=2$,
the critical density is the geometric mean $\rho_{\rm crit}\sim
\sqrt{\rho_{\rm Pl}\rho_*}$ and we have $\dpl\sim\delta_{\rm hol}$.)


\subsection{Slow-roll parameters}

For later convenience, we define the first three slow-roll parameters 
as
\ba
\epsilon &\equiv& 1-\frac{\cH'}{\cH^2}\,,\\
\eta &\equiv& 1-\frac{\vp''}{\cH\vp'}\,,\\
\xi^2 &\equiv& \frac{1}{\cH^2}\left(\frac{\vp''}{\vp'}\right)'+ \epsilon+\eta-1\,,
\ea
which coincide with the standard definitions in synchronous time
\[
\epsilon\equiv-\frac{\dot H}{H^2}\,,\qquad \eta\equiv -\frac{\ddot\vp}{H\dot\vp}\,,\qquad \xi^2\equiv \frac{1}{H^2}\left(\frac{\ddot\vp}{\dot\vp}\right)^.\,.
\]
The parameter $\epsilon$ will be especially important later on and we can rewrite it as
\begin{eqnarray}
  \epsilon     &=& 4\pi G\frac{\a}{\nu}\frac{\vp'^2}{\cH^2}\left(1-\frac13\frac{\rmd\ln \nu}{\rmd\ln p}\right)-\frac{\rmd\ln\a}{\rmd\ln p}\nonumber\\
         &=& 4\pi G\frac{\vp'^2}{\cH^2}\left\{1+\left[\a_0+\nu_0\left(\frac{\s}{6}-1\right)\right]\dpl\right\}+\frac{\s\a_0}{2}\dpl\,,\label{edp}
\end{eqnarray}
using the Raychaudhuri equation (\ref{Ray}) in the first step.  Notice
that the symbol $=$ in the last line of equation \Eq{edp} implicitly hides
the ${\rm O}(\dpl)$ truncation. This note of caution applies to any of the
equations below, where ${\rm O}(\dpl^2)$ terms are dropped as required for
self-consistency of perturbed equations. In contrast, the slow-roll
approximation will always be invoked explicitly and indicated with the
symbol $\approx$.

The derivatives of $\epsilon$ and $\eta$ are
\ba
\epsilon' &=& 2\cH\left(\epsilon+\frac{\rmd\ln\a}{\rmd\ln p}\right)\left[\epsilon-\eta+\frac{\rmd\ln\a}{\rmd\ln p}-\frac{\rmd\ln\nu}{\rmd\ln p}-\frac13\frac{\rmd^2\ln\nu}{\rmd\ln p^2}\left(1-\frac13 \frac{\rmd\ln\nu}{\rmd\ln p}\right)^{-1}\right]\nonumber\\
&&-2\cH\frac{\rmd^2\ln\a}{\rmd\ln p^2}\nonumber\\
&=& 2\cH\epsilon(\epsilon-\eta)-\s\cH\ve \dpl\,,\\
\eta' &=& \cH(\epsilon\eta-\xi^2)\,,
\ea
where
\be\label{vare}
\ve\equiv \a_0\left(\frac{\s}{2}+2\e-\eta\right)+\nu_0\left(\frac{\s}6-1\right)\e\,. 
\ee
While in standard inflation $\epsilon$ is almost constant whenever it
is small (since the classical part of $\epsilon'$ is quadratic in the
parameters), depending on the size of the quantum correction $\s\ve\dpl$
the quantity $\epsilon'$ could be of the same order as
$\epsilon$. However, we expect $\dpl$ to be small in the typical
setting.

For a given background $a(\tau)$ and $\vp(\tau)$,
the slow-roll parameters are functionally identical to the classical
case. Clearly, the potential required to give rise to such an
 evolution is different, as one can see also from the (later useful)
relations 
\ba
V_{,\vp}    &=& \frac{\cH \vp'}{\nu p}\left(\eta-3+2\frac{\rmd\ln\nu}{\rmd\ln p}\right)\nonumber\\
&=& \frac{\cH \vp'}{\nu p}\left(\eta-3-\s\nu_0\dpl\right)\,,\label{Vp}\\
V_{,\vp\vp} &=& \frac{\cH^2}{\nu p}\left[3(\epsilon+\eta)-\eta^2-\xi^2+2(3-\epsilon-2\eta)\frac{\rmd\ln\nu}{\rmd\ln p}+4\frac{\rmd^2\ln\nu}{\rmd\ln p^2}-4\left(\frac{\rmd\ln\nu}{\rmd\ln p}\right)^2\right]\nonumber\\
&=& \frac{1}{\nu p}\left(-m_\vp^2+\cH^2\s\mu_\vp\dpl\right)\,,\label{Vpp}
\ea
where
\ba
m_\vp^2 &\equiv& \cH^2[\eta^2+\xi^2-3(\epsilon+\eta)]\,,\label{mvp}\\
\mu_\vp &\equiv& \nu_0\left(\s-3+\epsilon+2\eta\right)\,.
\ea


\section{Background solutions}\label{baso}

Let $\phi$ be a set of generic scalar variables and let us write the background equations of motion, as well as the soon-to-be-found Mukhanov equation for the scalar perturbation, as ${\cal O}[\phi]=0$, where ${\cal O}$ is a (possibly non-linear) differential operator. One can drop quantum terms of order higher than $\dpl$ and split each variable into a classical part $\phi_\c$ and a quantum correction $\phi_\q\dpl$ \cite{CH},
\be \label{phi}
\phi= \phi_\c+\phi_\q\dpl\,,
\ee
so that each equation becomes 
\be
{\cal O}_\c[\phi_\c]+\left\{{\cal O}_\c[\phi_\q]+{\cal O}_\q[\phi_\c]\right\}\dpl=0\,.
\ee
Requiring that the classical and quantum part vanish separately (a condition which defines what is meant by $\phi_\c$) yields two equations:
\be
{\cal O}_\c[\phi_\c]=0\,,\qquad {\cal O}_\c[\phi_\q]+{\cal O}_\q[\phi_\c]=0\,.\label{lange}
\ee
This splitting strongly resembles the one into coarse- and
fine-grained perturbations in stochastic inflation
\cite{OLM,SB2,YiV1,YiV2}; in fact, for a Klein--Gordon scalar the second equation \Eq{lange} is nothing but a `complementary' Langevin-type equation for a quantum
variable with a noise term sourced by the classical part.

For example, consider the scalar field and scale factor profiles
\be
\vp= \vp_\c+\vp_\q\dpl\,,\qquad a= a_\c+a_\q\dpl\,.
\ee
The Hubble parameter can be written as
\be
\cH=\cH_\c+\cH_\q\dpl\,,\qquad \cH_\q=\frac{a_\q}{a_\c}\left[\frac{a_\q'}{a_\q}-(1+\s)\cH_\c\right]\,.
\ee
Also, the scalar potential $V$ is expanded in a Taylor series around $\vp_\c$,
\ba
V(\vp) &=& V(\vp_\c)+V_{,\vp}(\vp_\c)\vp_\q\dpl\equiv V_\c+V_\q\dpl\,,\label{Vvp}\\
V_{,\vp}(\vp) &=& V_{,\vp}(\vp_\c)+V_{,\vp\vp}(\vp_\c)\vp_\q\dpl\equiv V_{,\vp\,\c}+V_{,\vp\,\q}\dpl\,.\label{Vvpvp}
\ea
Plugging these expressions into the Friedmann and Klein--Gordon equations \Eq{frw} and \Eq{kg}, we obtain a pair of classical equations,
\ba
\cH_\c^2 &=& \frac{8\pi G}{3}\left(\frac{{\vp_\c'}^2}{2}+a_\c^2V_\c\right)\,,\\
0 &=& \vp_\c''+2\cH_\c\vp_\c'+a_\c^2 V_{,\vp\,\c}\,,
\ea
plus another pair of relations involving the correction functions $\vp_\q$ and $a_\q$:
\ba
\cH_\c\cH_\q&=&\frac{4\pi G}{3}\left[\vp_\c'(\vp_\q'-\s\cH_\c\vp_\q)+\frac{\a_0-\nu_0}{2}\,{\vp_\c'}^2+a_\c^2(\a_0V_\c+V_\q)+2a_\q V_\c\right],\label{frwq}\\
0&=&
\vp_\q''+2\cH_\c(1-\s)\vp_\q'+\s\cH_\c^2(\s+\e_\c-3)\vp_\q+a_\c^2(\nu_0V_{,\vp\,\c}+V_{,\vp\,\q})\nonumber\\
&&+2a_\q V_{,\vp\,\c}+\vp_\c'(2\cH_\q+\s\nu_0\cH_\c)\,.\label{kgq}
\ea

At this point we look for special background solutions with exactly constant slow-roll parameter $\e$, i.e., with a scale factor expanding as a power-law:
\be\label{powl}
a=a_\c=|\tau|^n\,,\qquad n\leq-1\,,
\ee
where $\tau<0$ and the limit $n\sim -1$ corresponds to de Sitter
spacetime. By this definition, the quantum corrections $a_\q$
and $\cH_\q$ vanish identically.
At the classical level, one gets the power-law solution \cite{LM}
\ba
\vp_\c &=& \vp_0 \ln|\tau| =\pm\sqrt{\frac{n(n+1)}{4\pi G}} \ln|\tau|\,,\label{vp0}\\
V   &=& V_0\, \rme^{-2(n+1)\vp/\vp_0}=\frac{n(2n-1)}{8\pi G}\, \rme^{-2(n+1)\vp/\vp_0}\,,
\ea
for which $V_\c=V(\vp_\c)=V_0|\tau|^{-2n-2}$ and
\be\label{eex}
\e=\eta_\c=\xi_\c=1+\frac1n.
\ee
Let us see if there exist solutions of the form $\vp_\q=\vp_{\q 0}
|\tau|^b$, distinguishing two cases:
\begin{itemize}
\item If $n\neq -1$,
\be\label{Vq}
V_\q = -2(n+1)V_0\frac{\vp_{\q 0}}{\vp_0}|\tau|^{b-2n-2}\,,
\ee
for $V_\q$ as defined in (\ref{Vvp}). Equations \Eq{frwq} and \Eq{kgq} become
\ba
0 &=& \vp_0\vp_{\q 0}\left[(b-\s n)-2(n+1)\frac{V_0}{\vp_0^2}\right]|\tau|^{b-2}+\vp_0^2\left[\frac{\a_0-\nu_0}{2}+\a_0\frac{V_0}{\vp_0^2}\right]|\tau|^{-2}\,,\nonumber\\\label{frwq2}\\
0&=&
\vp_{\q 0}\left[b(b-1)+2bn(1-\s)+\s n(\s n+1-2n)+4(n+1)^2\frac{V_0}{\vp_0^2}\right]|\tau|^{b-2}\nonumber\\
&&+\nu_0\vp_0\left[\s n-2(n+1)\frac{V_0}{\vp_0^2}\right]|\tau|^{-2}\,.\label{kgq2}
\ea
It turns out that, if $\vp_{\q 0}\neq 0$, the solution requires $b=0$ and the equalities
\ba
\frac{\vp_{\q 0}}{\vp_0} &=& \left(\frac{3n}{1+n}\a_0-\nu_0\right)\frac{1}{2(\s n+2n-1)}=\left(\nu_0-\frac{3}{\e}\,\a_0\right)\frac{1-\e}{2(3+\s-\e)}\,,\nonumber\\\label{vpq0}\\
0&=&\nu_0\left[6(3-\e)-\s(3+\s-\e)\right]-3\a_0\left[2(3-\e)-\frac{\s}{\e}(3-\s-\e)\right]\,.\label{bo}
\ea
The potential then reads
\be
V=\left[1-2(n+1)\frac{\vp_{\q 0}}{\vp_0}\dpl\right] V_\c\,,
\ee
while the second and first slow-roll parameters are
\ba
\eta  &=& \eta_\c-\s^2n\frac{\vp_{\q 0}}{\vp_0}\dpl\,,\\
\xi^2 &=& \xi_\c^2-\s^2(\s n+1+n)\frac{\vp_{\q 0}}{\vp_0}\dpl\,.
\ea
An exact solution is
\be\label{can}
\s=0\,,\qquad \a_0=\nu_0\,,\qquad V=(1-\a_0\dpl)V_\c\,,
\ee
while for $\s\gtrsim {\rm O}(1)$ and small $\e$ the expression \Eq{bo} is satisfied if $\a_0$ and $\nu_0$ obey
\be\label{bohh}
\nu_0\left[18-\s(3+\s)\right]-3\a_0\left[6-\frac{\s}{\e}(3-\s)\right]\approx 0\,.
\ee
For $0<\s<3$, the solution has $\a_0,\nu_0>0$ if $\e>\s(3-\s)/6$ (e.g., $\e>1/3$ if $\s=1$ or $\s=2$). Because of the lower bound on $\e$, this solution prefers the limiting values $\s\sim0$, $\s\sim 3$ if extreme slow roll is to be realized.

When $\s\geq3$, $\a_0$ and $\nu_0$ have opposite sign and $|\nu_0|\gg|\a_0|$. This case is
excluded by the above considerations on inverse-volume operators, which require $\a_0$ and $\nu_0$ to be both positive.

If $\vp_{\q 0}=0$, equations \Eq{frwq2} and \Eq{kgq2} are solved for
\be\label{lass}
\a_0=\frac{\e}{3}\,\nu_0\,,\qquad \s=3-\e\,,\qquad V=V_\c\,.
\ee
This solution has $\nu_0\gg\a_0$ and $2<\s<3$. For a given
  background, $\e$ is constant per (\ref{eex}); a relation to the
  constant $\alpha_0$, $\nu_0$ and $\sigma$ may thus be
  acceptable. However, the required tuning of general quantization
  parameters to background parameters makes this solution very special.

By construction and consistently, all these power-law solutions obey
equation \Eq{edp}. Their qualitative features are summarized in table
\ref{table1}. 
\item The last exact power-law case we consider is de Sitter, $n=-1$,
  whose classical solution is $\vp_\c={\rm const}$. There we cannot
  use equation \Eq{Vq} because the potential (\ref{Vq}) was derived
  using (\ref{vp0}) which is ill-defined for $n=-1$. From equation \Eq{frwq} we
  simply obtain $V_\q=-\a_0V_\c$. Since
  $V_\q=V_{,\vp}(\vp_c)\vp_\q$, this implies that $\vp_q$ is constant.
  But $\dpl(a)$ is not constant, so that $\vp$ is not constant and $V(\vp)$ can be reconstructed from the evolution. Combining (\ref{phi}) and (\ref{Vvp}), we find
\be
V(\vp)=V_\c(1-\a_0\dpl)= V_\c+V_{,\vp\,\c}(\vp-\vp_\c)\,.
\ee
Equation \Eq{kgq} becomes $\s(\s-3)\tau^{-2}\vp_{\q 0}+a^2m^2\vp_{\q 0}=0$,
where $m^2\equiv V_{,\vp\vp}(\vp_\c)$. The
reconstructed potential above is linear in $\vp$, so $m^2=0$.
If $\vp_{\q 0}\neq 0$, this implies that either $\s=0$ or $\s=3$.
\end{itemize}

\TABLE[ht]{
\caption{\label{table1} Inflationary power-law solutions $a=|\tau|^n$ with $0<\e=1+1/n<1$ and exponential potential. $\vp_\c(t)$ and $\vp_{\q 0}$ are given by equations \Eq{vp0} and \Eq{vpq0}, respectively.}
\begin{tabular}{|c|c|c|}\hline
$\vp$                      & $\s$      & $\a_0,\nu_0$ \\ \hline
													 & $\s=0$    & $\a_0=\nu_0$ \\
$\vp_\c(t)+\vp_{\q 0}\dpl$ & $0<\s<3$  & $\a_0,\nu_0>0$\, if\, $\e>\frac{\s(3-\s)}6$ \\
                           & $\s\geq3$ & ${\rm sgn}(\a_0)=-{\rm sgn}(\nu_0)$,\,\, $|\nu_0|\gg|\a_0|$ \\\hline
$\vp_\c(t)$  & $2<\s\lesssim 3$ & $\a_0=\frac{\e}{3}\,\nu_0\ll \nu_0$ \\ \hline
\end{tabular}
}


In the next section we will find that not all of these solutions
will be compatible with a certain consistency relation on quantum
counterterms. We have not shown that the above solutions are
attractors in configuration space, a necessary condition for adopting
them as valid backgrounds. In the next section we will assume this is
the case, since in the quasi-classical regime the dynamics is very
close to general relativity. Anyway, the structure of perturbation
equations and observables does not change qualitatively if one expands
about a more general quasi-de Sitter solution.

Also, setting $a=a_\c$ as in equation \Eq{powl} will not result in any loss
of generality in solving the Mukhanov equation. Since the coefficients
in the second equation \Eq{lange} will depend only on $\cH_\c$, the
structure of the quantum corrections in the solution will be always
the same, regardless of $\cH_\q$. However, here we see a possible
drawback of assuming an exact power-law expansion, equation \Eq{powl}:
these background solutions constrain the range of $\s$, $\a_0$, and
$\nu_0$, and from this analysis it is not obvious whether more
general, quasi-power-law backgrounds will admit a different parameter
space. We will leave also this question to future investigations. For
the time being, we show that there exist quasi-power-law expansions as
exact solutions and we briefly sketch a profile corresponding to a
perturbed de Sitter background, $a_\c=-1/\tau=\cH_\c$ and $a_\q\neq
0$. Assuming that $\vp_\q=\vp_{\q 0}$, $V_\q={\rm const}$,
equation \Eq{kgq} yields, as before, either $\vp_{\q 0}=0$ or
$m^2=\s(3-\s)$, while equation \Eq{frwq} becomes 
\be
a_\q'+\left(\frac{1+\s}\tau-\frac{8\pi G}{3}V_\c\right) a_\q-\frac{4\pi G}{3}\frac{\a_0V_\c+V_\q}{\tau^2}\equiv a_\q'+\left(\frac{1+\s}\tau-b_1\right) a_\q-\frac{b_2}{\tau^2}=0\,.
\ee
The solution is
\be
a_\q(t)= \frac{\rme^{b_1\tau}}{\tau}\left[\frac{a_0}{\tau^\s}-b_2 E_{1-\s}(b_1\tau)\right]\,,
\ee
where $E$ is the exponential integral function. For $b_1>0$ and integer $\s>0$,
\be\nonumber
a_\q=\frac{a_0\rme^{b_1\tau}+{\rm Pol}[{\rm O}(\tau^{\s-1})]}{\tau^{\s+1}}\,.
\ee
The last term, a polynomial of degree $\s-1$, dominates at early times ($\tau\to-\infty$) and $a_\q\sim |\tau|^{-2}$, while at late times ($\tau\to 0$) one has $a_\q\sim |\tau|^{-1-\s}$.

When $\s=0$, at early times $E_1(b_2\tau)\sim \rme^{-b_2\tau}/(b_2\tau)$ and $a_\q\sim |\tau|^{-2}$, while at late times $E_1(b_2\tau)\sim-\ln|\tau|$ and $a_\q\sim \ln|\tau|/|\tau|$.


\section{Scalar perturbations}\label{sper}


\subsection{Counterterms}

As mentioned in the introduction, the corrected perturbed equations
feature counterterms proportional to $\dpl$ in addition to the primary
correction functions $\alpha$ and $\nu$, which guarantee consistency
of the constraint algebra at ${\rm O}(\dpl)$ order. Consistency in a given
scheme then uniquely relates the counterterms, which all vanish
classically, to the primary correction functions, but also restricts
the range of parameters in $\alpha$ and $\nu$. Before using the
counterterms in perturbation equations, we evaluate these consistency
conditions in relation with table~\ref{table1}.  In the following, we
shall rewrite the counterterms and equations of motion of \cite{BHKS2}
according to the $\dpl$-expansion. To keep notation light, background
quantities will not be denoted with bars as in \cite{BHKS2}. Also,
contrary to this reference we shall expand all intermediate
expressions to linear order in counterterms (for instance,
$(1+f)(1+h)= 1+f+h+{\rm O}(\dpl^2)$, and so on). Explicitly, the counterterms are
\ba
f   &=& \frac{1}{\s}\frac{\rmd\ln\a}{\rmd\ln p}\nonumber\\
    &=& -\frac{\a_0}{2}\dpl\,,\\
f_1 &=& f-\frac13\frac{\rmd\ln\nu}{\rmd\ln p}\nonumber\\
    &=& \frac12\left(\frac{\s\nu_0}{3}-\a_0\right)\dpl, \label{f1}\\
h  &=& 2\frac{\rmd\ln\a}{\rmd\ln p}-f\nonumber\\
    &=& \a_0\left(\frac12-\s\right)\dpl\,,
\ea
and
\ba
g_1 &=& \frac13\frac{\rmd\ln\a}{\rmd\ln p}-\frac{\rmd\ln\nu}{\rmd\ln p}+\frac29 \frac{\rmd^2\ln\nu}{\rmd\ln p^2}\label{g1}\\
    &=& \frac{\s}2\left(\frac{\s\nu_0}{9}+\nu_0-\frac{\a_0}{3}\right)\dpl,\nonumber\\
f_3 &=& f_1-g_1\label{f3}\\
    &=& \frac12\left[\a_0\left(\frac{\s}3-1\right)-\frac{2\s\nu_0}{3}\left(\frac{\s}6+1\right)\right]\dpl\,.\nonumber
\ea
There is also the extra consistency condition
\be\nonumber
2\frac{\rmd f_3}{\rmd\ln p}+3(f_3-f)=0\,,
\ee
which makes some of the parameters dependent:
\be\label{extra}
\a_0\left(\frac{\s}6-1\right)-\nu_0\left(\frac{\s}6+1\right)\left(\frac{\s}3-1\right)=0\,,
\ee
so that
\ba
g_1 &=& \left[\nu_0\left(\frac{\s}{3}+1\right)-\a_0\right]\dpl\,,\\
f_3 &=& \left[\frac{\a_0}{2}-\nu_0\left(\frac{\s}6+1\right)\right]\dpl\\
    &=&\frac{1}{2}\frac{3\alpha_0}{\s-3} \dpl\,,\label{f3ter}
\ea
the last expression being valid only if $\s\neq 3$.

It is interesting to notice that, for the second but not last time,
the minisuperspace and first lattice parametrization
\Eq{ci}--\Eq{nulc} show an incompatibility with independent results:
in these parametrizations, equation \Eq{extra} is never respected.
However, with the new range \Eq{newc} for $\s$, the equation can
easily be satisfied by the solutions of table \ref{table1}. Let us
compare case by case with equation \Eq{extra}. The solution in the first
line of the table (equation \Eq{can}) is also an exact solution of
\Eq{extra}. This is the limiting case of the solution in the second
line (equation \Eq{bohh}) for $\s\ll 1$, giving $\a_0\approx \nu_0$; for
general values $\s<3$ (e.g., $x=0$ with $m<3$ or $x=-1/4$ with $m<6$)
solutions exist in this class for positive $\a_0$ and $\nu_0$, while
$\a_0=0$ when $\s=3$. The solutions in the third line of table
\ref{table1} are already excluded by the general constraint
$\a_0,\nu_0>0$.\footnote{Even ignoring the constraint on the
sign, these solutions would be inconsistent or trivial. For $3<\s<6$,
both \Eq{bohh} and \Eq{extra} require $\a_0$ and $\nu_0$ to have
opposite sign, but while $|\nu_0|\gg|\a_0|$ for \Eq{bohh},
equation \Eq{extra} asks them to be of about the same magnitude. When
$\s=6$, $a_0=0=\nu_0$. If $\s>6$, equation \Eq{extra} requires $\a_0$ and
$\nu_0$ to have the same sign, in contrast with \Eq{bohh}; so again
$a_0=0=\nu_0$.} Finally, the last solution in the table, equation \Eq{lass}
combined with \Eq{extra}, is non-trivial and inflationary only if
$\s=3$, but this collapses to de Sitter, $\e=0$. To summarize, our
solutions will span the range 
\be\label{finalrange} 
0\leq \s\leq 3\,,
\ee 
with preference to the extremum values if $\a_0$ and $\nu_0$ are
positive. In fact, $\e\gtrsim 1/3$ is not very small for 
$1\lesssim\s\lesssim2$, which 
should lead to unviable deviations from scale invariance.


\subsection{Scalar perturbation equations}

An inhomogeneous perturbation $\delta\vp$ in the scalar field induces two gauge-invariant scalar modes $\Phi$ and $\Psi$ in the metric, which are proportional to each other \cite{BHKS2}:
\be\label{PhPs}
\Phi=(1+h)\Psi\,.
\ee
After solving the equations for $\Psi$, the expression for $\Phi$ will be readily obtained via equation \Eq{PhPs}. The scalar-field perturbation and $\Psi$ are related by the diffeomorphism constraint (equation (90) of \cite{BHKS2})
\be\label{pdif}
4\pi G\frac{\a}{\nu}\vp'\delta\vp=\Psi'+(1+f+h)\cH\Psi\,.
\ee
Using this equation, one can show that the perturbed equation for $\Psi$ is\footnote{This is obtained by combining our equations \Eq{PhPs} and \Eq{pdif} with equation (82) of \cite{BHKS2}. In equation (82) one should correct the typographical error $\bar\a^2\Delta\Phi\to\bar\a^2\Delta\Psi$ \cite{Erratum}:
\be\label{usef}
\a^2\Delta\Psi-3\cH(1+f)[\Psi'+(1+f+h)\cH\Psi]=4\pi G\frac{\a}{\nu}(1+f_3)\left[\vp'\delta\vp'-\vp'^2(1+f_1+h)\Psi+\nu pV_{,\vp}\delta\vp\right]\,,
\ee
where $\Delta$ is the Laplacian in comoving spatial coordinates. Also, in equation (91) of \cite{BHKS2} one must replace $\bar\a^2(1+h)\Delta\Psi$ with $\bar\a^2\Delta\Psi$. Unfortunately, this typo propagated in some of the other equations, so our results supersede those of \cite{BHKS2} when in disagreement.}
\ba\label{psie}
\Psi''+\cH F\Psi'-\left(s^2\Delta+m_{\Psi}^2\right)\Psi=0.
\ea
The friction term is
\ba
F &=& 
2\left(1-\frac{\rmd\ln\a}{\rmd\ln p}\right)+(1+f+h)+3(1+f-f_3)-2\left(3-2\frac{\rmd\ln\nu}{\rmd\ln p}\right)+2\eta\nonumber\\
&=& 2\eta+\s F_0\dpl,
\ea
where
\be
F_0\equiv \nu_0\left(\frac{\s}6-1\right)-\frac{\a_0}2\,.
\ee
The (squared) propagation speed of the perturbation is
\be
s^2 = \a^2(1-f_3)= 1+\chi\dpl\,, 
\ee
where
\be\label{chipsi}
\chi\equiv \frac{\s\nu_0}{3}\left(\frac{\s}6+1\right)+\frac{\a_0}{2}\left(5-\frac{\s}3\right)\,.
\ee
Finally, the effective mass term is
\ba
m_{\Psi}^2 &=& \cH^2\left[2(\e-\eta)-2\frac{d f}{\rmd\ln p}+3\frac{\rmd\ln \a}{\rmd\ln p}-3(f-f_3)-4\frac{\rmd\ln \nu}{\rmd\ln p}\right.\nonumber\\
&&\left.+\e\left(\frac13\frac{\rmd\ln \nu}{\rmd\ln p}+f_1+f+2h\right)-2(f+h)\eta-\frac{h'}{\cH}\right]\nonumber\\
&=& \cH^2\left[2(\epsilon-\eta)-\s\mu_\Psi\dpl\right]\,,
\ea
where
\be
\mu_\Psi\equiv [2(\e-\eta)+(1+\s)]\a_0+\nu_0\left(\frac{\s}{6}-1\right)\,.
\ee
Taking equation (84) of \cite{BHKS2}, expanding it to leading order in quantum corrections, and making use of equations \Eq{Vp}, \Eq{Vpp}, and \Eq{PhPs}, one obtains the perturbed Klein--Gordon equation for the gauge-invariant perturbation $\delta\vp$:
\be\label{kgp}
\delta\vp''+2\cH B_1\delta\vp'-(s^2 \Delta-\nu pV_{,\vp\vp})\delta\vp-B_2\vp'\Psi'+2B_3\cH\vp'\Psi=0\,,
\ee
where
\bs\ba
B_1    &=& 1-\frac{\rmd\ln\nu}{\rmd\ln p}-\frac{\rmd g_1}{\rmd\ln p}\nonumber\\
&=& 1+B_{10}\dpl,\\
B_2 &=& 4+f_1+h+3g_1\nonumber\\
&=& 4+B_{20}\dpl,\\
B_3 &=& (1+f_1+h)\frac{\nu p V_{,\vp}}{\cH\vp'}-\frac{\rmd h}{\rmd\ln p}-\frac{\rmd f_3}{\rmd\ln p}\nonumber\\
&=& \eta-3 +B_{30}\dpl,
\ea
and
\ba
B_{10} &\equiv& \s\left[\nu_0\left(\frac{\s}{6}+1\right)-\frac{\a_0}2\right]\,,\\ 
B_{20} &\equiv& \frac{\s}{2}\left(\frac{\s\nu_0}{3}+\frac{10\nu_0}{3}-3\a_0\right)\,,\\
B_{30} &\equiv&  \s\left[\left(\frac{\nu_0}6-\a_0\right)\eta-\nu_0\left(\frac{\s}{12}+2\right)+\frac{\a_0}{2}(7-\s)\right]\,. 
\ea
\es

Before proceeding, we notice a potentially serious problem. In order to avoid superluminal propagation of signals, one should impose 
\be
s^2<\a^2\,,
\ee
where we used the fact that photons propagate with speed $\a$ greater than the classical one \cite{BH2}. Then, it should be $f_3>0$. 
For this to happen, we can have:
\begin{itemize}
\item $0\leq \s<3$: equation \Eq{f3ter} imposes $\a_0<0$.
\item $\s=3$: equations \Eq{f3} and \Eq{extra} impose, respectively, $\nu_0<0$ and $\a_0=0$.
\item $3<\s<6$: equation \Eq{extra} imposes $\a_0$ and $\nu_0$ to have opposite sign.
\item $\s=6$: equations \Eq{f3ter} and \Eq{extra} impose, respectively, $\a_0>0$ and $\nu_0=0$; this case is allowed.
\item $\s>6$: this case, too, is allowed, with both $\a_0$ and $\nu_0$ strictly positive.
\end{itemize}
Unfortunately, for non-negative $\a_0$ and $\nu_0$, $f_3$ is
negative unless $\s$ be large enough, and this condition is hardly
compatible with inflation; see table~\ref{table1}. (In \cite{Erratum},
the values of parameters given for subluminal evolution correspond to
the case $\sigma>6$ here.)

To check whether superluminal propagation is an artifact of
linear perturbation theory or of the expansion in $\dpl$, one should
go beyond linear order in both expansions. The covariant formalism of
non-linear perturbation theory could be a useful tool for analyzing the
consistency of the effective constraint algebra. A possibility is that
holonomy corrections, which we have ignored, would play an important
role in this issue, which we shall put aside in this paper.  However,
even if this were the case in some regimes, one can always find
initial conditions so as to have dominant inverse-volume corrections;
thus, superluminal velocities might constitute a conceptual problem
with implications for the stability of the theory as a whole. On
the other hand, we note that inflationary models based on
superluminally propagating fields have been consistently formulated
\cite{MuV,BMV}. The case of superluminal motion found here therefore does not
necessarily mean a severe problem.

We reemphasize the importance of equation (\ref{extra}) and of the
  counterterms it comes from. It rules out the minisuperspace-related
  parametrizations and severely restricts the lattice one. In
  this way, consistency alone already subjects the theory to strict
  tests even before evaluating the phenomenology, to which we turn now.


\subsection{Conservation of curvature perturbation}\label{sper2}

The gauge-invariant linear comoving curvature perturbation is \cite{BHKS2}
\ba
\cR &=& \Psi+\frac{\cH}{\vp'}(1+f-f_1)\,\delta\vp\label{crcc}\\
    &=& \Psi+\frac{\cH}{\vp'}\left(1-\frac{\s\nu_0}{6}\dpl\right)\delta\vp\,.\label{crcc2}
\ea
In the absence of counterterms, conservation of the energy-momentum tensor implies that $\cR$ is constant at large scales \cite{WMLL}. One may ask if this result, which is not obvious in Hamiltonian formalism and for equation \Eq{crcc}, holds also in semi-classical LQC. To check it, 
we invert equation \Eq{usef} with respect to $\delta\vp'$ and employ \Eq{pdif}. Differentiating $\cR$ with respect to conformal time, we obtain
\be\nonumber
\cR'=(\a\nu+f-f_1-f_3)\frac{\cH}{4\pi G \vp'^2}\Delta\Psi+C\delta\vp\,,
\ee
where
\ba
C&=&4\pi G\frac{\a}{\nu}\vp'+\frac{\cH^2}{\vp'}\left[\frac{f'-f_1'}{\cH}-(1+f-f_1)\left(\e+2\frac{\rmd\ln\nu}{\rmd\ln p}+3f-3f_3\right)\right]\nonumber\\
&=&\frac{\cH^2}{\vp'}\left[\frac{f'-f_1'}{\cH}+\frac{\rmd\ln\a}{\rmd\ln p}+\left(\frac13\frac{\rmd\ln\nu}{\rmd\ln p}-f+f_1\right)\e-2\frac{\rmd\ln\nu}{\rmd\ln p}-3(f-f_3)\right]\nonumber\\
&=& 0\,.\label{conca}
\ea
Here, after using (\ref{edp}), the prefactor of the parameter $\e$ (which we are not assuming to be small) vanishes by virtue of equation \Eq{f1}, which also implies $(f'-f_1')/\cH= (2/3) \rmd^2\ln\nu/\rmd\ln p^2$. Together with $f-f_3=-(2/3)\rmd\ln\nu/\rmd\ln p+(1/3) \rmd\ln\alpha/\rmd\ln p+(2/9)\rmd^2\ln\nu/\rmd\ln p^2$ from (\ref{f3}) and (\ref{g1}), all terms are zero.
The resulting conservation of power is consistent with the picture of an effective perturbed FRW geometry that models the dynamics of a nearly isotropic universe in the presence of corrections from loop quantum gravity. Quantum-geometry corrections from this theory, with the perturbation equations used here, have been shown to deform not just the dynamics but also the underlying spacetime structure, inferred by an analysis of the algebra of constraints. The gauge transformations they generate no longer correspond to pure coordinate transformations because they do not obey strictly the classical algebra of spacetime deformations. As a consequence, classical results about the conservation of power may no longer apply. As seen here, the linear curvature perturbation is nevertheless conserved on large scales. This observation demonstrates that perturbations in the presence of quantum corrections can still be seen as those of an effective line element: The large-scale curvature perturbation of an FRW line element in conformal time amounts simply to a spatially constant rescaling of the scale factor, which should not be subject to non-trivial dynamics. By being conserved also in the presence of quantum corrections, the interpretation of the effective geometry as a line element is still meaningful. 

The result \Eq{conca} is due to the delicate cancellations between counterterms. Therefore,
\be
\boxed{\phantom{\Biggl(}\cR'=\left[1+\left(\frac{\a_0}{2}+2\nu_0\right)\dpl\right]\frac{\cH}{4\pi G \vp'^2}\Delta\Psi\phantom{\Biggl(}}\,,
\ee
and \emph{the curvature perturbation is conserved at large scales}.


\subsection{Mukhanov equation}\label{sper3}

Conservation of $\cR$ strongly suggests that one can write a simple Mukhanov equation in the variable 
\be\label{mukvar}
u=z\cR\,,
\ee
where $z$ is some background function. We can anticipate the main result with a very efficient trick, and then confirm it via a standard but tedious calculation. The trick is to notice that, at super-horizon scales, the comoving curvature perturbation is approximately constant, so that $u''\approx z''\cR$ and
\be\nonumber
u''-\frac{z''}{z}u \approx 0\,.
\ee
The objective now is to find this friction-free Mukhanov equation from the perturbed equations of motion. Start from equation \Eq{kgp} and choose for simplicity a spatially flat slice where $\Psi\approx0$. In order to remove the friction term, we need to define a field $u=a(1-\b\dpl)\delta\vp$, where $\b=B_{\rm 10}/\s$. Then,
\be\nonumber
\frac{u''}{1-\b\dpl}=\delta\vp''+2\cH(1+B_{\rm 10}\dpl)\delta\vp'+\dots\,.
\ee
Comparing with the Mukhanov variable \Eq{mukvar} and equation \Eq{crcc2}, one finds
\be\label{z}
z\equiv \frac{a\vp'}{\cH}\left[1+\left(\frac{\s\nu_0}{6}-\b\right)\dpl\right]= \frac{a\vp'}{\cH}\left[1+\left(\frac{\a_0}{2}-\nu_0\right)\dpl\right]\,.
\ee
The only missing term in the Mukhanov equation is the Laplacian, with coefficient $s^2$ as an inspection of equation \Eq{kgp} immediately shows. Thus we obtain
\be\label{muk}
u''-\left(s^2\Delta+\frac{z''}{z}\right)u =0\,,
\ee
a result valid exactly at all scales and at the linear perturbative level. The rigorous calculation begins with the Mukhanov variable \Eq{mukvar} and equation \Eq{z} with unknown $\b$. Differentiating $u$ twice, using equations \Eq{psie} and \Eq{kgp}, and using equation \Eq{pdif} to develop the $\Psi'$ term, we obtain equation \Eq{muk} plus just one extra term:
\be\nonumber
u''=\left(s^2\Delta+\frac{z''}{z}\right)u+2\s\cH\dpl\left[\nu_0\left(\frac{\s}{6}+1\right)-\frac{\a_0}2-\b\right]\left(\cH z\Psi-\eta\cH a\delta\vp-a\delta\vp'\right)+{\rm O}(\dpl^2)\,.
\ee
The extra term vanishes if $\b$ is chosen as above.

It may seem that equation (\ref{muk}) is not covariant since only the
spatial-derivative term is corrected. However, despite appearance this
is not the case: The quantum-corrected equations of motion correspond
to a deformed algebra of constraints as found in \cite{BHKS}, and the
constraints determine what form gauge transformations take. In general
relativity, the gauge transformations are spacetime diffeomorphisms
or changes of coordinates whose classical form implements the usual
notion of covariance. With corrected constraints obeying a deformed
algebra, the gauge transformations are not of the classical form, and
they do not correspond to the usual notion of coordinate changes. Even
though the underlying structure of a `quantum manifold' (perhaps
non-commutative) on which the modified transformations could be
interpreted as simple coordinate changes is unknown, the (generalized)
covariance of (\ref{muk}) under these deformed transformations is
guaranteed by the derivation of the equations of motion used here from
an anomaly-free set of constraints.

It is quite remarkable that scalar perturbations are ultimately
governed by such a simple equation as \Eq{muk}. However, the existence
of one Mukhanov variable obeying one equation in closed form is not
unexpected, as it could have been inferred by using the
Hamilton--Jacobi method for constrained Hamiltonian systems developed
in \cite{GNR,Lan94}. In particular, the reduced phase space
obtained after solving the constraints and factoring out their gauge
flows has one local degree of freedom, parametrized by the curvature
perturbation and its conjugate momentum. There must be a closed form
for the dynamics on this reduced phase space, such that Hamiltonian
first-order equations of motion exist involving only ${\cal R}$ and
its momentum, and they are linear thanks to the linear perturbation
scheme used. As always, first-order Hamiltonian equations of motion
can be expressed as one second-order equation for the configuration
variable, here ${\cal R}$. The second-order equation in general may
have terms involving ${\cal R}''$, ${\cal R}'$, as well as ${\cal R}$,
which on large scales is an ordinary differential equation with
gradient-free coefficients (in momentum space, they are independent of the wave number $k$ defined below).
Thus, one may eliminate the last term involving ${\cal R}$ by substituting $y{\cal
R}$ for ${\cal R}$ for a suitable background function $y$, and a
constant mode for $y{\cal R}$ results. As a consequence, there must be
a conserved quantity $y{\cal R}$ whose existence can be seen without
any detailed calculations. Details are required to derive the form of
$y$, and the non-trivial result found here is that $y=1$.

From now on we expand linear perturbations in momentum space, a subscript $k$ indicating modes with comoving wavelength $2\pi/k$. The Laplacian becomes $\Delta\to-k^2$, and the Mukhanov equation
\be\label{mukk}
\boxed{\phantom{\Biggl(}u_k''+\left(s^2k^2-\frac{z''}{z}\right)u_k=0\phantom{\Biggl(}}\,.
\ee
The effective mass term is a combination of slow-roll parameters and quantum corrections. In fact,
\ba
\frac{z'}{z} &=& \cH(1+\e-\eta)+\s\left(\nu_0-\frac{\a_0}2\right)\cH\dpl\,,\\
\frac{z''}{z} &=& \cH^2\left(2+2\epsilon-3\eta-4\epsilon\eta+2\epsilon^2+\eta^2+\xi^2-\s\mu_u\dpl\right)\,,\\
\mu_u &\equiv&\frac{3\a_0}{2}+\nu_0(\s-3)+\left(\frac{5\a_0}{2}+\frac{\s\nu_0}{6}-2\nu_0\right)\e+2(\nu_0-\a_0)\eta\,.
\ea
When the slow-roll parameters are constant classically, as in any of the solutions of section \ref{baso}, one has
\be
\frac{z''}{z}=\frac{4\mu_1^2-1+4\mu_2\dpl}{4\tau^2}\,,
\ee
where 
\be
\mu_1=\frac12-n\,,\qquad \mu_2=\s n^2\left[\s(4n-\s n+1)\frac{\vp_{\q 0}}{\vp_0}-\mu_u\right]\,.
\ee
An exact solution of the Mukhanov equation does exist but it is too complicated and not very instructive. We proceed to solve this equation asymptotically. 

\subsection{Asymptotic solutions}

The moment of horizon crossing is, as usual, defined when the effective mass term equals the Laplacian term. Up to numerical factors, this happens when
\be
k|\tau|=1\,,
\ee
as in standard inflation. Super-horizon modes are characterized by $k|\tau|\ll 1$, while modes well inside the horizon have $k|\tau|\gg 1$. At large scales, we can ignore the $k^2$ term in equation \Eq{mukk}, so that
\be\label{ucz}
u_k\stackrel{k|\tau|\ll 1}{\sim} C(k) z\,,
\ee
where $C(k)$ is a normalization constant. To determine it, we must find the asymptotic behaviour of $u$ at small scales. There, one can ignore the mass term and consider the equation
\be\label{mukk2}
u_k''+(1+\chi\dpl)k^2u_k\approx 0\,.
\ee
Since all the analysis is valid only at first order in the quantum corrections, it is consistent to look for short-wavelength solutions of the form
\be\label{ansk}
u_{k\gg \cH}(\tau) = u_\c(k,\tau)[1+y(k,\tau)\dpl]\,,
\ee
where $u_\c$ is the solution of the classical Mukhanov equation and $y$ is some function. In particular, the only choice compatible with the Bunch--Davies vacuum in the infinite past is an incoming plane wave,
\be
u_\c=\frac{\rme^{-\rmi k\tau}}{\sqrt{2k}}\,.
\ee
The normalization here is the classical one, which one might
have to change for a vacuum matter state in a quantum geometry. In particular, the correction function $\nu$ multiplies the kinetic term of the scalar Hamiltonian, and thus affects the value of vacuum fluctuations. By the ansatz (\ref{ansk}), all these effects will be included once the equation of motion for $y$ is solved.

Plugging the ansatz \Eq{ansk} into \Eq{mukk2} we obtain an inhomogeneous equation for the function $y$:
\be\label{sig}
y''-2(\s\cH+\rmi k)y'+2\rmi k\s\cH y+\chi k^2=0\,,
\ee
where, for consistency, we have dropped the mass term $\s\cH^2(\s+\e-1)y$. At this point we expand $y$ in a power series,
\be\label{sumy}
y=\sum_{m=0}^{+\infty}y_m\tau^m\,,
\ee
and we pick a power-law background, $\cH=n/\tau$. Then, equation \Eq{sig} is
\ba
0&=& 2\s n(\rmi ky_0-y_1)\frac{1}{\tau}+[2\rmi k(\s n-1)y_1+2(1-2\s n)y_2+\chi k^2]\nonumber\\
 &&+\sum_{m=2}^{+\infty}[2\rmi k(\s n-m)y_m-(m+1)(2\s n-m)y_{m+1}]\tau^{m-1}\,.
\ea
These terms must vanish order by order separately. If $\s=0$, then $y_0$ is unconstrained, while $y_1=-\rmi(y_2/k+k\chi/2)$ and $y_m=2(2\rmi k)^{m-2}y_2/m!$ for all $m\geq2$. Summing the series, one obtains
\ba
y&=&y_0-\rmi\left(\frac{y_2}{k}+\frac{k\chi}{2}\right)\tau+\frac{y_2}{2k^2}\left(1+2\rmi k\tau-\rme^{2\rmi k\tau}\right)\nonumber\\
&=&\left(y_0+\frac{y_2}{2k^2}\right)-\frac{\rmi k\chi}{2}\tau-\frac{y_2}{2k^2}\rme^{2\rmi k\tau}\,.\nonumber
\ea
We can argue that $y_2=0$ because otherwise $u$ in equation (\ref{ansk})
would contain also an outgoing mode $\rme^{+\rmi k\tau}$. If $\s\neq 0$, one obtains the following conditions:
\ba
y_1 &=& \rmi ky_0\,,\\
y_2 &=& \frac{k^2}{2(2\s n-1)}[\chi-2(\s n-1)y_0]\,,\\
y_{m+1} &=& \frac{2\rmi k(\s n-m)}{(m+1)(2\s n-m)}y_m\,.
\ea
The recursive relation would determine the sum of the series \Eq{sumy}, but analytic continuation to the case $\s=0$ requires $y_m=0$ for $m\geq 2$. This fixes both $y_0$ and $y_1$ and the result is
\be
y=\frac{\chi}{2(\s n-1)}(1+\rmi k\tau)\,.
\ee
The normalization of equation \Eq{ucz} is thus obtained by imposing the junction condition $|u_{k\gg \cH}|=|u_{k\ll \cH}|$ at horizon crossing. Then,
\be\label{ukh}
|u_{k\ll \cH}|^2= \frac{1}{2k}\left[1+\frac{\chi}{\s n-1}\dpl(k)\right] \left[\frac{z}{z(k)}\right]^2\,,
\ee
where $z(k)=z(\tau=-1/k)$ and $\dpl(k)=\dpl(\tau=-1/k)\propto k^{n\s}$.


\subsection{Scalar spectrum, spectral index and running}\label{sper4}

The scalar spectrum is defined as the two-point correlation function of the curvature perturbation $\cR$ over a momentum ensemble at large scales, evaluated at horizon crossing:
\be
{\cal P}_{\rm s} \equiv \frac{k^3}{2\pi^2z^2} \left\langle |u_{k\ll\cH}|^2\right\rangle\Big|_{k|\tau|=1}\,.
\ee
The scalar spectral index is defined as
\be
n_{\rm s}-1 \equiv \frac{\rmd\ln {\cal P}_{\rm s}}{\rmd\ln k}\,.
\ee
For a power-law background, we have
\ba
{\cal P}_{\rm s}(k) &=& \frac{G}{\pi}k^{2(1+n)}\left[1+\left(\frac{\chi}{\s n-1}-\a_0+2\nu_0+2\s n\frac{\vp_{\q 0}}{\vp_0}\right)\dpl\right]\,,\\
n_{\rm s}-1 &=& 2(1+n)+\s n\left(\frac{\chi}{\s n-1}-\a_0+2\nu_0+2\s n\frac{\vp_{\q 0}}{\vp_0}\right)\dpl\,.
\ea
We can obtain more portable expressions by writing the spectrum on a
general quasi-de Sitter background. Since, using (\ref{edp}),
\be\nonumber
z^2=\frac{a^2}{4\pi G}\left\{\e-\left[\nu_0\left(\frac{\s}{6}+1\right)\e+\frac{\s\a_0}{2}\right]\dpl\right\}\,,
\ee
we get
\be\label{scasp}
\boxed{\phantom{\Biggl(}{\cal P}_{\rm s} = \frac{G}{\pi}\frac{\cH^2}{a^2\e}\left(1+\g_{\rm s}\dpl\right)\phantom{\Biggl(}}\,,
\ee
where we used $k=\cH$ and
\be \label{gs}
\g_{\rm s}\equiv \nu_0\left(\frac{\s}{6}+1\right)+\frac{\s\a_0}{2\e}-\frac{\chi}{\s+1}\,.
\ee
Notice that if $\s=0$, the quantum correction is constant and the only change with respect to the classical case is the normalization of the spectrum. In that case, $\g_{\rm s}=\nu_0-5\a_0/2$ could be of either sign. If $\g_{\rm s}\neq 0$, there is a large-scale enhancement of power because $\dpl\sim a^{-\s}\sim (1/|\tau|)^{-\s}\sim k^{-\s}$ at horizon crossing. The magnitude of the effect depends on the value of $\g_{\rm s}$ but we notice that, if $\s\neq 0$, $\g_{\rm s}\sim 3\a_0/(2\e)$ unless $\nu_0\sim\a_0/\e\gg\a_0$. Therefore, one could obtain a sizable enhancement unless $\a_0\lesssim {\rm O}(\e)$ (which is the case, typically).
Since this enhancement is of potential interest for comparisons
  with observations, we trace back where the inverse of the slow-roll
  parameter in the expression for $\gamma_{\rm s}$ came from. It arises due
  to the $\epsilon$-independent term in $z^2$ above, which in turn is
  a direct consequence of the presence of gravity corrections in the
  Raychaudhuri equation (\ref{Ray}), as opposed to just stress-energy
  modifications. As with several other key phenomena pointed
  out here, this feature is a consequence of corrections to the
  structure of spacetime geometry: corrections in the terms ${\cal
  H}'$, ${\cal H}^2$ of the Raychaudhuri equation (or the isotropic
  Einstein tensor) can be obtained only by changing the geometrical
  form of gravity.

Momentum derivatives are converted into conformal time derivatives via 
\be\nonumber
\frac{\rmd}{\rmd\ln k}\approx\frac{1}{\cH}\frac{\rmd}{\rmd\tau}\,,
\ee
so that the scalar index is
\be
\boxed{\phantom{\Biggl(}n_{\rm s}-1 = 2\eta-4\e+\s\g_{n_{\rm s}}\dpl\phantom{\Biggl(}}\,,\label{ns}
\ee
where
\be
\g_{n_{\rm s}} \equiv \frac{\ve}{\e}-\a_0\left(1-\frac{\eta}{\e}\right)-\g_{\rm s} = \a_0-2\nu_0+\frac{\chi}{\s+1}\,.
\ee
Since the quantum correction is small, the scalar index does not deviate too much from scale invariance. If $\s=0$, there are no corrections at all. If $\s\neq 0$, the sign of the correction depends on the choice of the parameters in the parameter space. We have seen that power-law/quasi de Sitter solutions have $\s\lesssim 3$, so it is immediate to associate scale invariance with small values of $\s$. The naturalness of this range is further stressed in the concluding section by an independent argument.

Interestingly, the running of the spectral index is dominated by the
quantum correction (unless $\s=0$):
\ba
\a_{\rm s}&\equiv& \frac{\rmd n_{\rm s}}{\rmd\ln k}\\
          &=&2(5\e\eta-4\e^2-\xi^2)+\s(4\ve-\s \g_{n_{\rm s}})\dpl \sim \dpl\,.
\ea
This result signals a qualitative departure from classical inflation, since the quantum correction may be larger than ${\rm O}(\epsilon^2)$. The details will depend on the chosen background, as the slow-roll parameter themselves can contain quantum corrections.


\section{Tensor perturbations}\label{tper}

The linearized equation of motion for tensor modes has been computed in \cite{BH2} and solved in \cite{CH} for quasi-classical inverse volume corrections. In this section we review and improve these results, eventually obtaining the cosmological consistency relations.


\subsection{Mukhanov equation}

When only inverse-volume corrections are taken into account and in the absence of anisotropic stress, the equation of motion for the individual tensor mode $h_k$ is \cite{BH2}
\be\label{mue}
h_k''+2\cH\left(1-\frac{\rmd\ln\a}{\rmd\ln p}\right) h_k'+\a^2k^2 h_k=0\,.
\ee
Defining
\be
w_k\equiv \tilde a h_k\,,\qquad \tilde a\equiv a\left(1-\frac{\a_0}{2}\dpl\right)\,,
\ee
we get the Mukhanov equation
\be\label{mukkh}
\boxed{\phantom{\Biggl(}w_k''+\left(\a^2k^2-\frac{\tilde a''}{\tilde a}\right)w_k=0\phantom{\Biggl(}}\,,
\ee
where
\ba
\frac{\tilde a'}{\tilde a} &=& \cH\left(1+\frac{\s\a_0}2\dpl\right)\,,\\
\frac{\tilde a''}{\tilde a} &=& \cH^2\left[2-\e+(3-\s-\e)\frac{\s\a_0}2\dpl\right]\,.
\ea
Equation \Eq{mukkh} is formally identical to the scalar Mukhanov equation and the analysis is exactly the same up to the substitutions
\be\nonumber
z\to \tilde a\,,\qquad \chi\to2\a_0\,.
\ee
The final result is the analogue of equation \Eq{ukh},
\be\label{ukh2}
|w_{k\ll \cH}|^2= \frac{1}{2k}\left[1+\frac{2\a_0}{\s n-1}\dpl(k)\right] \left[\frac{\tilde a}{\tilde a(k)}\right]^2\,.
\ee


\subsection{Tensor spectrum, spectral index and running}

The tensor spectrum is
\be
{\cal P}_{\rm t} \equiv \frac{32 G}{\pi}\frac{k^3}{\tilde a^2} \left\langle |w_{k\ll\cH}|^2\right\rangle\big|_{k|\tau|=1}\,,
\ee
so that in de Sitter ($n=-1$)
\be
\boxed{\phantom{\Biggl(}{\cal P}_{\rm t} \equiv \frac{16G}{\pi}\frac{\cH^2}{a^2} 
\left(1+\g_{\rm t}\dpl\right)\phantom{\Biggl(}}\,,\label{tesp}
\ee
where
\be
\g_{\rm t}\equiv \frac{\s-1}{\s+1}\a_0\,.
\ee
As for the scalar spectrum, barring special values of the parameters ($\g_{\rm t}=0$ when $\s=1$ or $\a_0=0$) there is a power enhancement at large scales because of $\dpl\sim k^{-\s}$, albeit the prefactor might not be as large as in equation \Eq{gs}. This type of enhancement has been seen in earlier numerical studies of the LQC tensor power spectrum, but it is difficult to exploit it observationally due to limitations by cosmic variance.

The tensor index and its running are
\be
n_{\rm t}\equiv\frac{\rmd\ln {\cal P}_{\rm t}}{\rmd\ln k}\,,\qquad \a_{\rm t}\equiv \frac{\rmd n_{\rm t}}{\rmd\ln k}\,,
\ee
so that
\be
\boxed{\phantom{\Biggl(}n_{\rm t}= -2\e-\s\g_{\rm t}\dpl\phantom{\Biggl(}}\,,\label{nt}
\ee
and
\be
\a_{\rm t}=-4\e(\e-\eta)+\s(2\ve+\s\g_{\rm t})\dpl\,.
\ee


\subsection{Tensor-to-scalar ratio}

The last piece of information we want to extract is the tensor-to-scalar ratio
\be
r\equiv \frac{{\cal P}_{\rm t}}{{\cal P}_{\rm s}}\,.
\ee
From equations \Eq{scasp} and \Eq{tesp} one obtains
\be\label{tts}
r = 16\e[1+(\g_{\rm t}-\g_{\rm s})\dpl]\,,
\ee
which yields the consistency relation
\be\label{tts2}
\boxed{\phantom{\Biggl(}r = -8\{n_{\rm t}+[n_{\rm t}(\g_{\rm t}-\g_{\rm s})+\s\g_{\rm t}]\dpl\}\phantom{\Biggl(}}\,.
\ee
Here we implicitly assumed that $\g_{\rm s}$ is not too large, so that the expansion in $\dpl$ is still meaningful. In quasi de Sitter regime $\e\ll 1$, so that $\g_{\rm s}={\rm O}(1)$ if $\s\a_0\sim {\rm O}(\e)$. This means that either $\s$ or $\a_0$ or both should be small.

Unless $\s=0$ or $\s=1$ (for which $\gamma_{\rm t}=0$), the tensor-to-scalar ratio is no longer proportional to the tensor index. Detection of a non-zero $r$ would require either a consistent deviation from de Sitter in standard cosmology or a sufficiently large quantum correction in de Sitter LQC.

As already explained for (\ref{muk}), equations (\ref{mukk}) and
(\ref{mukkh}) are covariant under the deformed transformations
generated by the anomaly-free set of corrected constraints. The
deformation gives rise to a new type of quantum effects which could
not be present for higher-curvature effective actions usually expected
of quantum gravity; it is possible only thanks to quantum corrections
to the geometry of space or even the manifold structure. The Mukhanov
equations for scalar and tensor modes are not only corrected, they also
acquire corrections of different forms. The scalar equations has a
correction given by $s^2=\alpha^2(1-f_3)$, while the tensor equation
is corrected by $\alpha^2$. The counterterm $f_3$ cannot typically
be set to zero, and so the corrections for scalar and tensor modes
differ. This difference, in turn, makes possible changes to the
tensor-to-scalar ratio which may provide a key signature of loop
quantum gravity.


\section{Discussion}\label{disc}

To summarize the main results of this paper:
\begin{itemize}
\item Different parametrizations of LQC inverse-volume quantum corrections are presented. A particular case is the so-called mini-superspace parametrization. Lattice refinement parametrizations are introduced in a self-contained way reorganizing a number of related results in the literature.
\item Cosmological solutions with power-law scale factor are found for the LQC background equations of motion in the presence of small inverse-volume corrections $\dpl$. Their form depends on the parameter range, they generalize the unique classical power-law solution with exponential potential and are first order in the quantum corrections. These solutions are further restricted by an anomaly cancellation condition and the requirement of slow roll.
\item For the first time, the already-known equations of motion for scalar perturbations with inverse-volume corrections are consistently rewritten as first-order expressions in $\dpl$ and combined into a single Mukhanov equation. The LQC comoving curvature perturbation is shown to be conserved at large scales, just as its classical counterpart.
\item The Mukhanov equation is solved asymptotically and the scalar spectrum and index are constructed.
\item The already-known LQC Mukhanov equation for tensor perturbations is conveniently rewritten and solved using the same techniques. Tensor observables are extracted and combined with scalar observables, thus providing the full set of inflationary observables in linear perturbation theory.
\end{itemize}

As long as the slow-roll approximation is valid, the structure of the
cosmological observables is valid for any background,
although the coefficients of the quantum corrections themselves do
depend on the background. In this final section we discuss how
they can be used to restrict models of loop quantum cosmology, making
the framework falsifiable. Details will be provided in separate
publications \cite{BCT}. For such an endeavor, it is crucial to obtain
independent information on the main correction parameter $\dpl$ and on
different versions of the parametrization. For instance, as seen in
section \ref{s:Estim}, a combination with holonomy corrections is
interesting and shows a powerful interplay between these main two
types of quantum-geometry corrections.

If inflation is assumed, the minisuperspace parametrization is under
tight pressure: neither consistent power-law background solutions nor
a nearly scale-free spectrum can be found in that case (unless $\dpl$
be very small; see below). On the other hand, the minisuperspace
parametrization may still be viable if an alternative scenario of
structure formation can be found. In this context, holonomy
corrections are of particular interest not only by providing an
additional consistency condition in combination with inverse-volume
corrections, but also because they can easily trigger bounces at least
in isotropic models whose matter energy is dominated by the kinetic
term. (In general, it has not been shown that isotropic bounces occur
as a natural consequence of holonomy corrections.) It would thus be of
interest to develop linear perturbation equations around those models
and analyze the structure evolution through the bounce, or perhaps new
scenarios providing the generation of structure during a phase before
(not after) the big bang. However, compared to inverse-triad
corrections such ideas are currently hampered by several major
difficulties: (i) Holonomy corrections have so far not been
implemented in consistent deformations of linear perturbation
equations. (ii) Strong quantum-geometry corrections are required to
evolve through the bounce; no expansion in parameters such as $\dpl$
used here could be done. (iii) There are several indications as to the
strong sensitivity of evolution through the bounce to initial
conditions of perturbations \cite{InhomThroughBounce} or even the
quantum state \cite{BeforeBB}, discussed in the context of cosmic
forgetfulness.

The lattice parametrization is consistent with inflation and
holonomy corrections at the homogeneous level and could yield
strong effects according to section \ref{s:Estim} because $\s$
can be small, but that again depends on the details of $\dpl$. We have
seen that $\dpl$ is an eigenfunction of the operator $\rmd/\rmd\ln k$ with
eigenvalue $-\s$, so observables of higher order in the slow-roll
parameters (e.g., the index running) are corrected by a term which is
always of the form ${\rm O}(\s^n)\dpl$: it is first-order in $\dpl$ and
$n$-th order in $\s$. If $\s={\rm O}(1)$, this quantum correction is equally
important at any slow-roll order, if not increasing with the
order. This situation does not seem natural inasmuch as it would imply
that higher-order $k$ derivatives of the inflationary spectra are all
on the same footing. Then, the notion that the spectra can be
approximated by a power law would have to be abandoned. On the other
hand, if $\s\ll 1$ the quantum correction is suppressed by higher and
higher powers of $\s$, so there is a sort of balancing effect which
keeps ${\rm O}(\s^n)\dpl$ small at all orders in the slow-roll
parameters. This leads to the speculation that small values of $\s$
are more sensible, because for large $\s$ quantum corrections would
dominate in higher-order observables.

With $\s\ll1$ preferred, inverse-volume corrections are of the form
$1+cp^{-\s/2}$ with a small exponent $\s$. They affect not only the
equations for an expanding universe but also the dispersion relations
of waves propagating in a quantum spacetime. Corrections for these
equations are more difficult to derive because the situation is not as
symmetric as the one of perturbations of an isotropic spacetime. But
if they turned out to be of a similar form ($\dpl$ with small exponent
$\s$) for a variable related to the particle's energy, severe
observational pressure could be put on loop quantum gravity by a
combination of cosmological and astroparticle observations.

The first-order cosmological observables already give a wealth of information about the early universe but it is natural to ask oneself what happens at second order, e.g., when looking at the bispectrum and possible non-Gaussian signatures. Posed in LQC, this question is less harmless than in classical general relativity. In fact, when going to higher orders in perturbation theory one would get more parameters initially, because there are more options for 
counterterms. The counterterms would then be fixed by consistency conditions. This would not necessarily add extra conditions for the parameters arising at lower orders, but something like this might happen. We do not know yet whether loop quantum gravity as a whole is consistent, so at some point parameters might be overconstrained. If that were the case, at least in LQC, one would have to understand if a non-perturbative cancellation of anomalies (which one did not see at the perturbative level) takes place. Therefore, an extension of our results to a second-order analysis would be most welcome not only for the purpose of finding the bispectrum, but also in order to further check the self-consistency of the theory.

Before we conclude, we emphasize that we have considered in
detail only one type of corrections (inverse-volume) and no
complete set of effective equations implementing all the effects
expected from loop quantum gravity. Even so, the conclusions we draw
are reliable because they point out characteristic phenomena from the
corrections considered. An elimination of these effects by including
other phenomena (most importantly, those due to holonomies) can be
expected only under very fine-tuned conditions. The equations provided
here can thus be used to place bounds on the free parameters of loop
quantum gravity, and to rule out some parametrizations as extensively
done in this paper. In all these cases, we see the importance of
sufficiently general parametrizations for models to be able to stand
up to phenomenological pressure; conceptual preferences or `natural'
choices of parameters may not always be the ones that survive
stringent analysis. As our examples demonstrate, it is important to
combine constraints from different sources. For instance, the
minisuperspace parametrization is ruled out by (i) the consistency
condition provided by anomaly-freedom, (ii) the interplay of
inverse-volume with holonomy corrections, and (iii) the
phenomenological requirement of a nearly scale-invariant spectrum. A single
inconsistency could always be evaded by questioning the condition
violated, especially in a situation in which no tight derivations from
an underlying full theory exist. But as inconsistencies pile up,
models should eventually be dropped. In this way, models of loop
quantum gravity and loop quantum cosmology are already falsifiable not
just by internal consistency considerations but also by comparison
with observations.

Our results are interesting also because they highlighted a number of issues which are definitely worthy of further attention:
\begin{itemize}
\item The consistency of the slow-roll background solutions, of the
anomaly cancellations, and of the physical observables only with
respect to the second lattice parametrization urges us to study the
latter in greater detail. Such an endeavor requires a better
understanding of the full theory and its reduction to perturbations
around isotropic models.
\item Perturbations can propagate with superluminal speed. Either this is
an artifact of linear perturbation theory, is curable
with a particular choice of the parameters, or may give rise to
  non-standard inflationary scenarios as in \cite{BMV}. The minisuperspace
parametrization is safe as $f_3>0$ in that case (large $\sigma>3$),
while the lattice refinement parametrization with $\s\ll 1$ requires
$\a_0\leq 0$. This observation may be the one putting the most
severe constraints on the lattice refinement parametrization, while
the minisuperspace parametrization is under much stronger pressure
from other consistency conditions.
\end{itemize}
We believe that addressing these points and the mutual tension between different parametrizations and the physical viability of the perturbations will stimulate the advance in the field.


\begin{acknowledgments}
We thank G M Hossain and M Kagan for useful discussions, and
S Tsujikawa for a careful reading of the draft. This work was
supported in part by NSF grant 0748336.
\end{acknowledgments}


\end{document}